# Single-Photon Counting CMOS Detectors for the Habitable Worlds Observatory

**Edwin Alexani,[a*] Justin P. Gallagher,[a] Donald F. Figer[a]**

[a]Center for Detectors, Rochester Institute of Technology, Rochester, NY, USA

Abstract: The Habitable Worlds Observatory (HWO) is NASA's current New Great Observatory concept. HWO will observe the universe to search for life beyond Earth. It will need unprecedented instrument stability and detector performance. However, no single detector technology currently satisfies HWO's demanding specifications. Complementary metal-oxide-semiconductor (CMOS) single-photon counting detectors (SPCDs) are excellent candidates for further development, as they already satisfy several specifications for HWO. We report the results of a characterization program for 9.4 Mpixel CMOS SPCDs. We measure a dark current of 0.0005 $e^-$/s/pixel, a read noise of 0.19 $e^-$/pixel, a peak quantum efficiency of 88% at 485 nm, negligible crosstalk and persistence, and we present conversion gain measurements. The detector satisfies several key HWO performance requirements, including deep sub-electron read noise and low dark current, while additional development is required in ultraviolet quantum efficiency and full mission qualification. We include a simulated low-flux observation using characterization results, along with a roadmap for developing CMOS SPCDs to satisfy HWO specifications, especially at UV wavelengths.

## 1. Introduction

A major goal of astrophysics is to search for life on exoplanets. The 2020 Decadal Survey supports this effort and recommends the development of a "future large Infrared/Optical/Ultraviolet telescope optimized for observing habitable exoplanets and general astrophysics" [1]. The Survey notes that achieving these scientific goals will require "a period of mission and technology maturation." As such, the Habitable Worlds Observatory (HWO) is NASA's current mission concept to achieve these goals. The HWO Technology Development Plan identifies technology gaps that must be addressed for mission success [2]. The HWO Technology Roadmap (hereafter HWO Roadmap) designates candidate technologies for further development [3].

Detector performance is a threshold technology gap for HWO [2]. Single-photon counting detectors (SPCDs) will enable sensitive measurements for future strategic astrophysics missions such as HWO. Complementary metal-oxide-semiconductor (CMOS) SPCDs have low dark current, low read noise, high quantum efficiency, radiation tolerance, on-chip digitization, and per-pixel readout [4] [5] [6] [7] [8]. They exist in formats from 1 to 163 Mpixels, with pixel sizes ranging from 1 to 8 μm [9] [10] [11] [12] [13] [14]. Consequently, the HWO Roadmap baselines CMOS SPCDs as a candidate detector technology to further develop for HWO [3]. This paper presents a comprehensive laboratory characterization of the 9.4 Mpixel Fairchild Imaging (FI) HWK4123 single-photon counting CMOS detector and evaluates its suitability for HWO. We report measurements of dark current, read noise, conversion gain, floating diffusion capacitance, linearity, quantum efficiency, crosstalk, persistence, and glow, compare the results with HWO requirements, and identify the remaining technology gaps.

As part of a long-term development plan, our team is characterizing the HWK4123 CMOS SPCD. Table 1 lists detector characteristics reported by FI [9]. The HWK4123 is increasingly relevant in astrophysics, as it is deployed in commercially available camera systems used in laboratory and astronomical testbeds, and is selected for the future Lazuli Space Observatory [15] [16] [17] [18] [19] [20]. In fact, our team is currently reducing observing data for the open star cluster M36 acquired with one such camera (QUEST) at the C.E.K. Mees Observatory (Figure 1), and has imaged supernova SN 2025rbs in NGC 7331.

*Corresponding author: Edwin Alexani, ea2191@rit.edu

| Characteristic | Value |
|---|---|
| Format [pixels] | 4096 × 2304 |
| DC [$e^{-}$/s/pixel] | 0.01 at 253 K |
| Peak QE | 90% |
| Pixel size [μm × μm] | 4.6 × 4.6 |
| RN [$e^{-}$ RMS] | 0.25 at 5 fps |
| Programmable gain | 1×, 8×, 16×, 32× |

Table 1. This table lists HWK4123 characteristics reported by FI [9].

CMOS SPCD pixels have a low-capacitance sense node. This increases the voltage response of the sense node to a single charge carrier, *i.e.*, the conversion gain (V/$e^{-}$). As such, a single charge carrier produces a voltage response that far exceeds the read noise of the detector, making CMOS SPCDs single-photon counting and photon-number-resolving [5] [21] [22] [23]. Some CMOS SPCDs, such as the HWK4123, have an on-chip programmable gain amplifier (PGA). The PGA allows the user to adjust the effective system conversion gain (ADU/$e^{-}$), enabling a trade-off between noise performance and dynamic range. A larger gain amplifies the analog (V) signal from the pixel before it reaches the readout electronics, increasing the resolution of the signal per charge carrier at the cost of a smaller dynamic range (see Table 4).

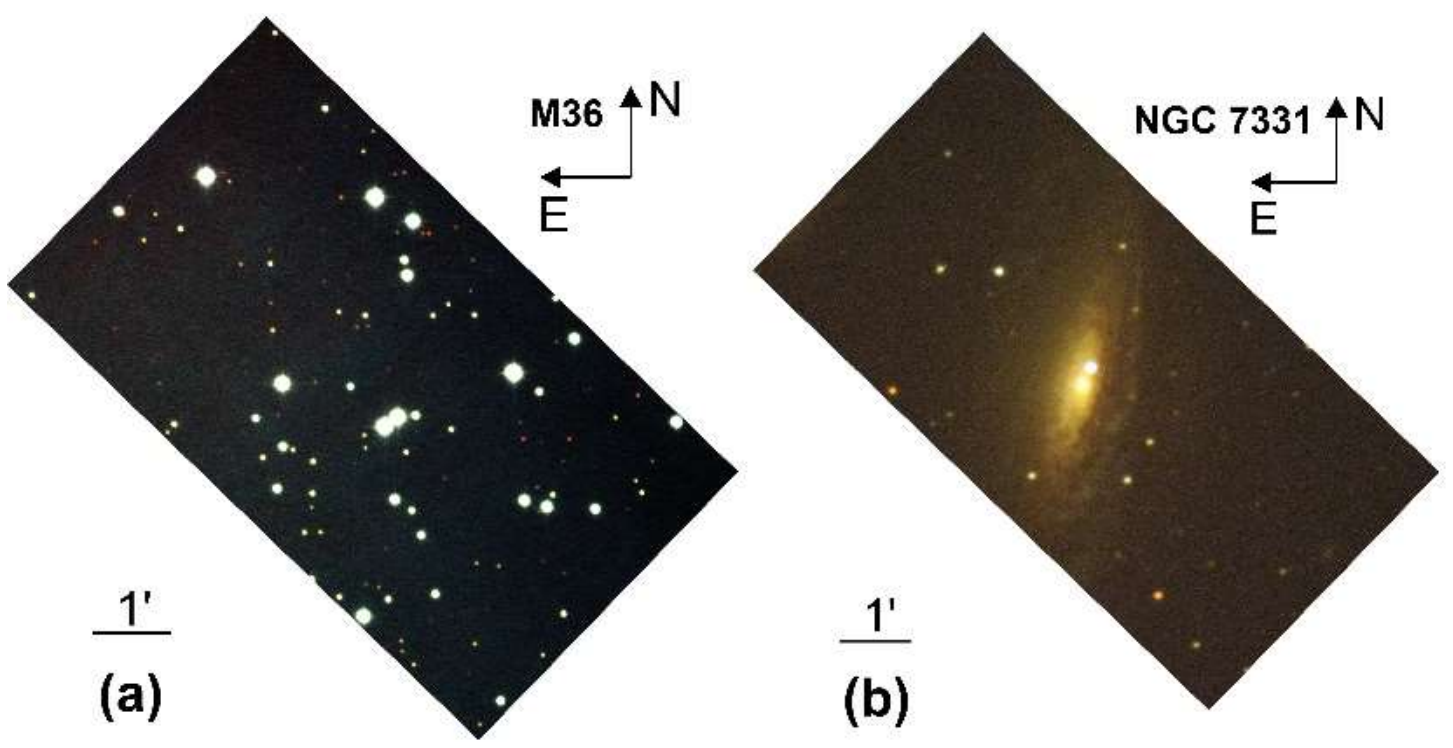

Figure 1. The panels show false-color images of *(a)* the open star cluster M36 and *(b)* the spiral galaxy NGC 7331. Blue, green, and red correspond to frames acquired with the Optolong B, V, and R filters for M36, and U, V, and I filters for NGC 7331, where supernova SN 2025rbs appears bluish-white near the galactic core.

The photon counting histogram (PCH) in Figure 2 illustrates single-photon resolution for a single pixel in the HWK4123 detector. The PCH has a series of distinct features corresponding to resolved electron numbers. Each feature is an ensemble of signal samples that are assigned to a specific electron number but span a width in values expressed in analog-to-digital units (ADU). This width depends primarily on the read noise. In fact, read noise affects the bit error rate, which is the probability of assigning the wrong electron number to a measurement [24] [25]. The separation between the centers of these features depends on the system conversion gain (ADU/$e^{-}$). Therefore, a PCH shows resolved electron numbers for low read noise detectors with high conversion gain.

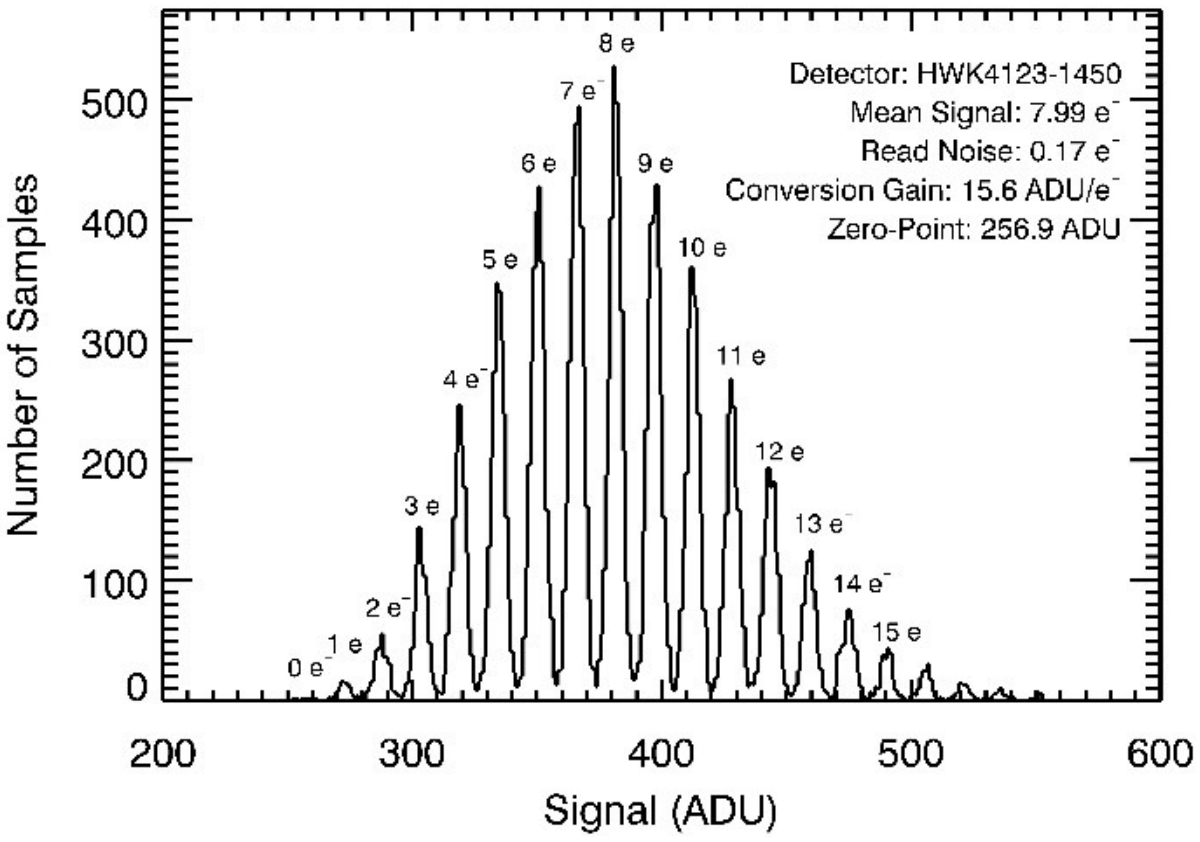

Figure 2. This plot is a photon counting histogram (PCH) for a single HWK4123 pixel based on 25,000 measurements. Each feature corresponds to a number of resolved electrons. The separation between adjacent features is the system conversion gain, which is 15.6 ADU/$e^{-}$.

The HWK4123 pixels have a five-transistor (5T) architecture [26], with the layout shown in Figure 3. Each pixel contains a pinned photodiode (PPD), a floating diffusion (FD) sense node, and five transistors: 1) the global reset gate, 2) the transfer gate (TG), 3) the reset gate (RG), 4) the source follower (SF), and 5) the row selection gate. Incoming photons generate electron-hole pairs in the silicon. The resulting photoelectrons accumulate in the potential well of the PPD. During readout, the TG changes this potential to transfer charges from the PPD to the FD. The sensor provides measurements using correlated double sampling (CDS). It measures the reset voltage of the FD, followed by the signal voltage. A single-slope analog-to-digital converter then compares and digitizes the output voltage.

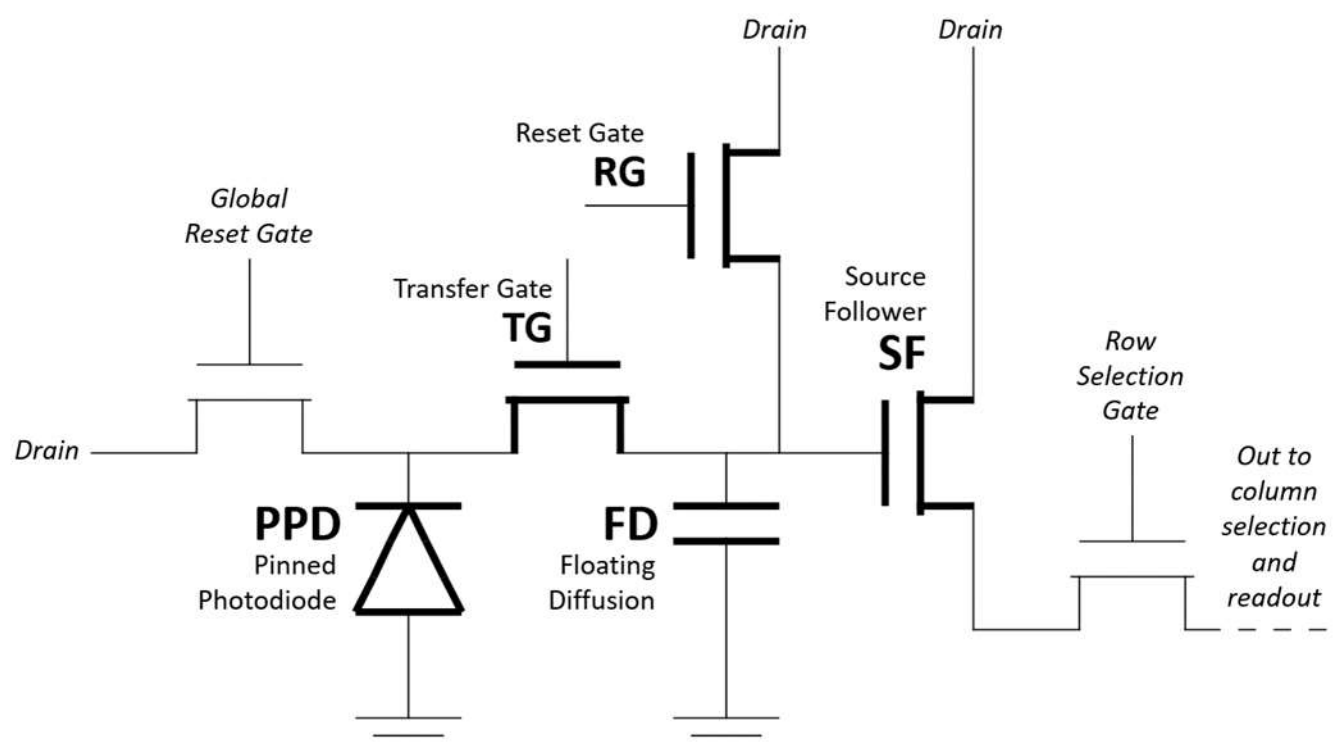

Figure 3. This diagram shows the architecture of a 5T pixel. The HWK4123 uses 5T pixels with a PPD. The bold components are discussed in the steps to calculate the FD capacitance.

## 2. Materials and Methods

This section describes the HWK4123 characterization program, including the hardware, software, and data analysis.

### *2.1. Characterization Experiments*

Detector characterization requires control of the detector, its environment, and supporting hardware setups. A light-tight, dark environment is required to measure dark current, read noise, crosstalk, and glow, while conversion gain, linearity, quantum efficiency, and persistence require controlled illumination. All measurements require temperature control, and all use correlated double sampling (CDS). The following sections describe the characterization experiments.

#### *2.1.1. Dark Current*

Dark current (DC) is the rate at which charge accumulates in a detector in the absence of light. Significant sources of DC include thermally-activated electron-hole pair generation and recombination, electron diffusion into the charge collection region, and generation at Si-$SiO_2$ interface states [27] [28] [29]. Thermal generation-recombination typically dominates DC and decreases with decreasing temperature [7]. DC charge carriers are indistinguishable from photoelectrons and contribute Poisson shot noise in quadrature to the total noise budget, thereby reducing the signal-to-noise ratio (SNR) of other measurements.

To measure DC, we acquire repeated destructive CDS dark exposures at different integration times. We subtract the frame with the shortest integration time from all subsequent frames to remove the digital offset and fixed bias structure, then average the frames at each integration time. We estimate DC from the slope of a linear fit to the dark signal as a function of time using both the median signal of each averaged frame and the signal per pixel. We fit the resulting pixel population DC histogram with a Gaussian to extract the mean of the distribution.

### *2.1.2. Read Noise and Bit Error Rate*

Read noise (RN) is the uncertainty in a pixel's measured signal. Stochastic processes during readout contribute to RN, including thermal Johnson-Nyquist noise, reset kTC noise, 1/f flicker noise, column amplifier noise, and analog-to-digital converter (ADC) quantization [30]. RN scales as the square root of the number of reads for independent, co-added frames. RN adds in quadrature to the total noise of a measurement and reduces SNR.

To measure RN, we acquire dark frames at a fixed integration time and calculate the standard deviation of each pixel's signal across frames. This metric represents the total noise, including DC shot noise. Consequently, we acquire dark frames at 346 μs and 2.5 fps, for which DC shot noise is negligible and RN dominates. We estimate RN per pixel as the standard deviation of its signal, produce an RN distribution, and report RN as the mean of a Gaussian fit to this distribution.

The bit error rate (BER) is the probability that a pixel reports an incorrect digital value during photon-counting operation. Read noise introduces a random fluctuation into the analog signal before digitization, which may alter the reported digital value. Consequently, BER can be expressed as a function of RN (Equation 34 of [24]). Several studies identify 0.3 $e^-$ as the RN threshold for single-photon counting [26] [31] [32], for which the BER is ~5%. BER decreases rapidly below this threshold, reaching ~0.6% at 0.2 $e^-$ and ~0.04% at 0.15 $e^-$.

### *2.1.3. Conversion Gain and the Capacitance of the Floating Diffusion*

During pixel readout, charge from the pinned photodiode (PPD) transfers to the floating diffusion (FD) sense node via the transfer gate (TG) and produces a voltage defined by the FD capacitance ($C_{FD}$). The FD connects to a source follower (SF) transistor that buffers the signal and couples the pixel to the column and row readout circuitry. The signal then propagates through transistors, amplifiers, and interconnects, each with finite resistance and capacitance. These components can modify the signal amplitude and therefore act as a gain stage in the system conversion gain ($CG_{sys}$).

The FD has a conversion gain of $CG_{FD}$ (V/$e^-$) defined by $C_{FD}$ and converts the collected charge into a voltage signal. The SF follows the FD signal and has a gain $G_{SF}$ (V/V), defined as the ratio of the SF output voltage to its input voltage. The SF connects the pixel to the remainder of the on-chip readout electronics, which we treat as a black box. The readout electronics convert the analog voltage into a digital value using the electronic conversion gain $CG_{elec}$ (V/ADU). $CG_{sys}$ is the convolution of these gains [33] and quantifies the response to a single charge carrier in units of ADU per electron (ADU/$e^-$), as shown in Eq. 1.

$$CG_{sys}(ADU/e^-) = CG_{FD}(V/e^-) \times G_{SF}(V/V) \times \left(CG_{elec}(V/ADU)\right)^{-1} \qquad \text{Eq. 1}$$

We can calculate $CG_{FD}$ using Eq. 1 and measurements for $CG_{sys}$, $G_{SF}$, and $CG_{elec}$:

$$CG_{FD}(V/e^-) = CG_{sys}(ADU/e^-) \times CG_{elec}(V/ADU) \times \left(G_{SF}(V/V)\right)^{-1} \qquad \text{Eq. 2}$$

We measure $CG_{sys}$ using the photon transfer experiment (PTE), which requires the acquisition of dark and illuminated frames over a range of integration times, up to detector saturation. We subtract dark current and bias from illuminated frames and compute the signal variance as a function of signal. The inverse slope of the linear region of this relation is the system conversion gain $CG_{sys}$, described by Equation 5.19 in [34], and reproduced here as Eq. 3:

$$\sigma^2_{READ+SHOT}(ADU) = \sigma^2_{READ}(ADU) + \frac{S(ADU)}{CG_{sys}(ADU/e^-)} \qquad \text{Eq. 3}$$

We have not yet accessed the SF and on-chip readout electronics to separately measure $G_{SF}$ and $CG_{elec}$. Instead, we leverage the HWK4123 pixel's use of correlated double sampling. The first sample measures the signal at the FD before charge transfer, while the second sample occurs after charge transfer. As such, we short the pixel reset gate (RG) to introduce a voltage change ($\Delta V$) between the first and second samples to produce a controlled change in the digital output signal ($\Delta ADU$). We define the measured conversion gain ($CG_{measured}$) as:

$$CG_{measured}(V/ADU) = \frac{\Delta V}{\Delta ADU} \quad \text{Eq. 4}$$

An external power source provides $\Delta V$ to the RG via an input-output (I/O) interface. The I/O interface has a coupling efficiency $\varepsilon_{coupling}$ (V/V) ranging from 0 to 1, and attenuates $\Delta V$ such that only $\varepsilon_{coupling} \times \Delta V$ reaches the RG. Therefore, we correct $CG_{measured}$ by $\varepsilon_{coupling}$ to determine the true voltage change introduced at the RG, and keep the associated $\Delta ADU$.

The RG shares a junction with the FD, and this junction lies upstream of the SF in the readout chain. Consequently, the product $CG_{measured} \times \varepsilon_{coupling}$ defines the $\Delta V$ from the same junction that the FD uses to connect to the SF, and records the corresponding $\Delta ADU$. This is the same definition as the product $CG_{elec} \times (G_{SF})^{-1}$, so we can write:

$$CG_{measured}(V/ADU) \times \varepsilon_{coupling}(V/V) = CG_{elec}(V/ADU) \times \left(G_{SF}(V/V)\right)^{-1} \quad \text{Eq. 5}$$

We substitute Eq. 5 into Eq. 2 to get:

$$CG_{FD}(V/e^{-}) = CG_{sys}(ADU/e^{-}) \times CG_{measured}(V/ADU) \times \varepsilon_{coupling}(V/V) \quad \text{Eq. 6}$$

We infer the value of $\varepsilon_{coupling}$ in Eq. 6 by replacing $CG_{FD}$ with the manufacturer's value (170 μV/e$^{-}$) [35] and using our measurements for $CG_{sys} \times CG_{measured}$ (~180 μV/e$^{-}$), giving $\varepsilon_{coupling} = 0.94$. We then calculate $CG_{elec}$ in Eq. 5 using $CG_{measured}$, $\varepsilon_{coupling}$, and $G_{SF}$. We assume $G_{SF} = 0.9$, consistent with reported values in the range 0.83–0.95 for CMOS detectors [36] [37].

Eq. 6 provides $CG_{FD}$, which is the inverse capacitance of the floating diffusion ($C_{FD}$) [33]. Using the capacitor equation ($C = Q/V$), we define $C_{FD}$ as:

$$C_{FD}\ (F) = \frac{1.6 \times 10^{-19}\ (Coul./e^{-})}{CG_{FD}\ (V/e^{-})} \quad \text{Eq. 7}$$

We report measurements for $CG_{sys}$ and $CG_{measured}$, and derived values for $CG_{elec}$ and $C_{FD}$ in Table 4, for all programmable gains listed in Table 1.

### *2.1.4. Linearity*

A detector is linear when equal changes in the input signal produce equal changes in the detector output, and is nonlinear when the relationship breaks down. Sources of nonlinearity include charge trapping, changes in floating-diffusion capacitance, source-follower gain variations, charge-collection effects near saturation, and nonideal behavior in readout electronics. Nonlinearity causes the detector response to deviate from the ideal linear response and introduces systematic measurement errors.

To measure linearity, we follow the methodology presented in [18] [38]. We use the PTE to acquire data spanning the detector's full dynamic range and identify the linear region. We fit a line to this region to define the ideal detector response. We then extrapolate the fit to the nonlinear response region and subtract the PTE data to produce residuals, which represent missing charge.

### *2.1.5. Quantum Efficiency*

Quantum efficiency is the ratio of the number of electrons (e$^{-}$) measured by the detector to the number of photons ($\gamma$) incident on the detector. We measure QE with the diode replacement method. We use a National Institute of Standards and Technology (NIST)-calibrated diode (DREF) to measure the photon flux at the output of an integrating sphere, and replace the diode with the device under test (DUT), *i.e.*, the HWK4123. A second NIST-calibrated diode (DMON) monitors the photon flux inside the sphere during the DREF and DUT measurements to correct for any changes in illumination.

We determine QE by comparing the DUT and DREF measurements. We subtract dark current and bias from DUT, DREF, and DMON measurements using corresponding dark frames. We correct for differences in illumination between DREF and DUT measurements using DMON, and scale by the ratio of the active area of an HWK4123 pixel to that of the DREF diode. We account for quantum yield (QY), which is the number of electrons collected per absorbed photon. For 400 to 1000 nm photons, we assume QY is unity [39] [40]. Table 2 lists the quantities used in Eq. 8 for the wavelength-dependent QE.

| Symbol | Quantity Description | Unit |
|---|---|---|
| $S_{DUT,on\text{-}off}$ | Dark-subtracted signal from DUT | $e^-$/pixel |
| $S_{REF,on\text{-}off}$ | Dark-subtracted signal from DREF | $\gamma$/diode |
| $S_{MON,DUT,on\text{-}off}$ | Dark-subtracted signal from DMON during DUT measurements | $\gamma$/diode |
| $S_{MON,REF,on\text{-}off}$ | Dark-subtracted signal from DMON during DREF measurements | $\gamma$/diode |
| $A_{DUT}$ | Area of one pixel from DUT | $m^2$/pixel |
| $A_{REF}$ | Active area of DREF | $m^2$/diode |
| QY | Quantum yield | dimensionless |
| QE | Quantum efficiency | $e^-/\gamma$ |

Table 2. This table lists the quantities used to determine the wavelength-dependent QE in Eq. 8.

$$QE(\lambda) = \frac{S_{DUT,on-off}}{S_{REF,on-off}} \times \frac{S_{MON,REF,on-off}}{S_{MON,DUT,on-off}} \times \frac{A_{REF}}{A_{DUT}} \times QY \quad \text{Eq. 8}$$

### *2.1.6. Crosstalk*

Crosstalk is the phenomenon where signal in one pixel changes the performance of a nearby pixel. Sources of crosstalk include interpixel capacitance between adjacent pixel sense nodes, charge diffusion of electrons in silicon prior to collection, and incomplete settling in the readout chain [41]. Crosstalk degrades SNR by redistributing signal and introducing correlated noise.

We represent crosstalk as the proportion of the signal in pixels surrounding a central pixel affected by a high-energy event (HEE). We define an HEE as a particle or photon interaction that produces a large number of charge carriers in a pixel. We use cosmic ray secondary particles and X-rays from a $^{55}Fe$ source to produce HEEs at a rate of ~70 events/s/cm$^2$. Some events may deposit charge across multiple pixels at once, leading to false positives, while others may saturate the central pixel and invalidate the measured crosstalk ratio. We select non-saturated pixels with normally and centrally incident interactions that produce a signal significantly larger than the signals in the surrounding pixels.

### *2.1.7. Persistence*

Persistence, also called lag, is the portion of a pixel's signal due to charge carriers produced by sources in previous images. We measured persistence by exposing the detector to a pulsed laser and monitoring the signal in subsequent dark frames. The pulse illuminates the detector, and the interval between pulses provides a dark environment. Any residual signal above the baseline dark current and read noise is attributable to persistence from the laser pulse.

### 2.1.8. Glow

Glow is the emission of photons by the detector during operation. Sources of glow include electroluminescence from pixel circuitry (unit-cell glow) during readout and from peripheral electronics such as multiplexers and output amplifiers (multiplexer glow). The emitted photons may be detected by the pixel, adding a background signal similar in effect to dark current.

We characterize glow with a spatial analysis of the DC population. Specifically, we compare DC in the imaging and reference pixels. Elevated DC in the reference pixels indicates glow.

## 2.2. Data Acquisition Software and Hardware

We developed and used software pipelines, written in the Interactive Data Language (IDL), to manage experiments, command electronics, and acquire and reduce data. The software controls all system components using parameter files that specify the detector analysis region and operating temperature, the number of frames and data cubes, the integration time per frame, time increments, and the illumination wavelength.

We characterize the HWK4123 using two complementary data acquisition systems. The first is a custom electronics platform, PHOTRON (Figure 4), that accommodates a single HWK4123 detector. The second is a commercial Hamamatsu Photonics ORCA Quest C15550-20UP camera system [15], hereafter QUEST (Figure 4). We have ten HWK4123 detectors that interface with PHOTRON and two QUEST cameras.

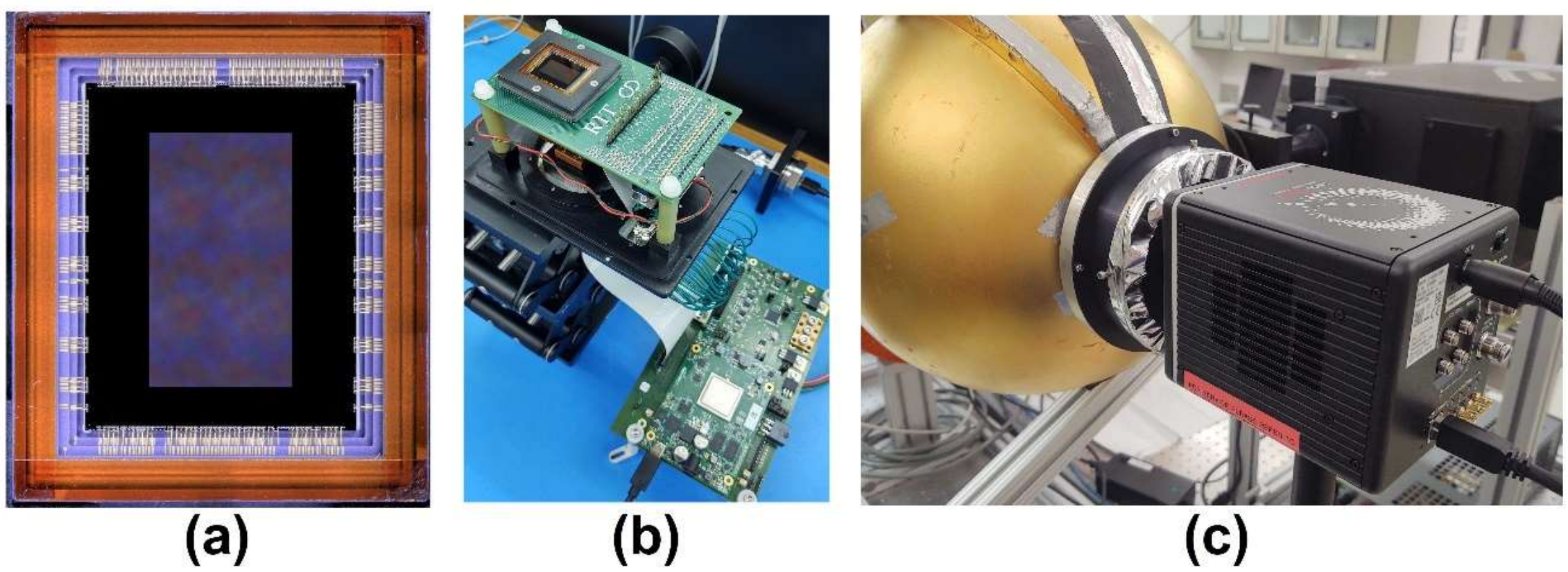

Figure 4. *(a)* The HWK4123 sensor has a standard ceramic land grid array (CGLA) package and measures 31.1 × 36.6 mm. The imaging pixel array is in the center of the sensor. *(b)* The PHOTRON system is shown outside the dewar. The HWK4123 mounts behind the black enclosure in the top-left corner. *(c)* The QUEST camera attaches to the output of an integrating sphere, and the sphere receives light from a monochromator.

We require a stable detector temperature to minimize variance in the measured signal, since signal sources such as DC depend on temperature. We meet this requirement in the PHOTRON system with a dewar, which provides a vacuum that prevents atmospheric contaminants from accumulating on the detector during low-temperature operation. The QUEST system cools the detector at atmospheric pressure using air cooling (253 K) or water cooling (233 K) [15].

Both systems use the same optical configuration from the light source to the integrating sphere output port. A quartz-tungsten-halogen lamp emits light into a monochromator equipped with filters, gratings, and mirrors to select the desired wavelength. We configure the monochromator output slit to provide a 4 nm bandpass that illuminates an integrating sphere, which has two output ports. We mount DMON to one of multiple ports on the integrating sphere to monitor the photon flux for all acquisitions. We attach the device under test (PHOTRON, QUEST, or DREF) to a second port.

For PHOTRON, the HWK4123 is mounted on a cold electronics board inside the dewar. The dewar has a calcium fluoride window with a flanged mount that couples to the integrating sphere output port and is secured with screws. For QUEST, the HWK4123 is in the camera head, which connects to the integrating sphere output port with optical tubes.

For persistence measurements, the optical system is replaced with a pulsed-laser configuration. A waveform generator controls the power supply of the 400–700 nm laser and defines the pulse timing. An optical fiber delivers the laser pulse to the QUEST system.

## 3. Laboratory Measurements Results

The HWK4123 detector characterization results include dark current, read noise, conversion gain, floating diffusion capacitance, linearity, quantum efficiency, persistence, crosstalk, and glow. The complementary acquisition systems have some differences. The PHOTRON system provides the user with greater control over temperature and programmable gain than the QUEST C15550-20UP camera. Unless explicitly stated, the measurements presented below derive from data acquired with PHOTRON at 32× gain, which provides photon-counting capability due to its 0.19 $e^-$ read noise. Any data presented for the QUEST use the dual-gain and the ultra-low light mode with lower read noise at 5 fps [15]. The QUEST system has higher QE than PHOTRON below 500 nm due to the antireflection-coated window in the QUEST package [15] [16] [35].

The characterization effort is ongoing; therefore, this report presents a representative subset of measurements. Table 3 summarizes the DC, RN, and QE results for the characterized sensors. The reported DC and RN values represent the mean of a Gaussian fit to the pixel population, and the uncertainty is the standard deviation (see Figure 5 and Figure 6 for examples). The QE measurements are not corrected for linearity. However, the QE data are in the linear regime of the detector, such that a linearity correction should modify the reported QE by no more than 5%. All measurements in Table 3 for PHOTRON are labeled with the sensor part number.

| | DC [$e^-$/s/pixel] | RN [$e^-$/pixel] | QE |
|---|---|---|---|
| QUEST-000488 | 0.0035 ± 0.0017 (233 K) | 0.21 ± 0.04 (233 K) | 88% (485 nm) (253 K) |
| QUEST-000621 | 0.0029 ± 0.0014 (233 K) | 0.21 ± 0.04 (233 K) | N/A |
| HWK4123-1441 | 0.0012 ± 0.0007 (232.6 K) | 0.18 ± 0.02 (200.5 K) | N/A |
| HWK4123-1450 | 0.00074 ± 0.00034 (218.5 K) | 0.19 ± 0.02 (223.4 K) | N/A |
| HWK4123-1442 | 0.00055 ± 0.00021 (208.9 K) | 0.19 ± 0.02 (267.1 K) | 84% (500 nm) (232 K) |
| HWK4123-1430 | 0.00053 ± 0.00021 (206.3 K) | 0.17 ± 0.02 (202.9 K) | 83% (500 nm) (280 K) |
| HWK4123-1453 | 0.00048 ± 0.00019 (204.1 K) | 0.19 ± 0.02 (215.5 K) | N/A |
| HWK4123-1427 | 0.00041 ± 0.00015 (184.5 K) | 0.19 ± 0.02 (224.7 K) | 84% (500 nm) (224 K) |

Table 3. This table lists DC, RN, and QE for HWK4123 sensors. Data taken with the QUEST are listed as QUEST, and those taken with PHOTRON are listed as HWK4123 with the corresponding part number.

### *3.1. Dark Current*

The dark current measurements are shown in Figure 5. The left panel shows the median signal of pixels in the analysis region, represented by the black dots. The red line is a fit to the dark signal as a function of time, with a slope of 0.00055 $e^-$/s/pixel. The right panel is a histogram of the DC evaluated for each pixel in the analysis region. The figure includes a cumulative curve that shows the histogram contains 99% (y-axis on the right) of all pixels in the analysis region. The red line is the mean value of a Gaussian fit to the distribution, giving a DC of 0.00048 ± 0.00019 $e^-$/s/pixel at 204.1 K. The DC measurements at different temperatures (Table 3) suggest the sensor may be approaching a DC plateau near 200 K, such that further cooling does not decrease DC as efficiently. The HWK4123 dark current satisfies the high-sensitivity UV/VIS instrumentation specification (0.002 $e^-$/s/pixel), but is five times larger than the visible coronagraph channel DC specification (0.0001 $e^-$/s/pixel) [3].

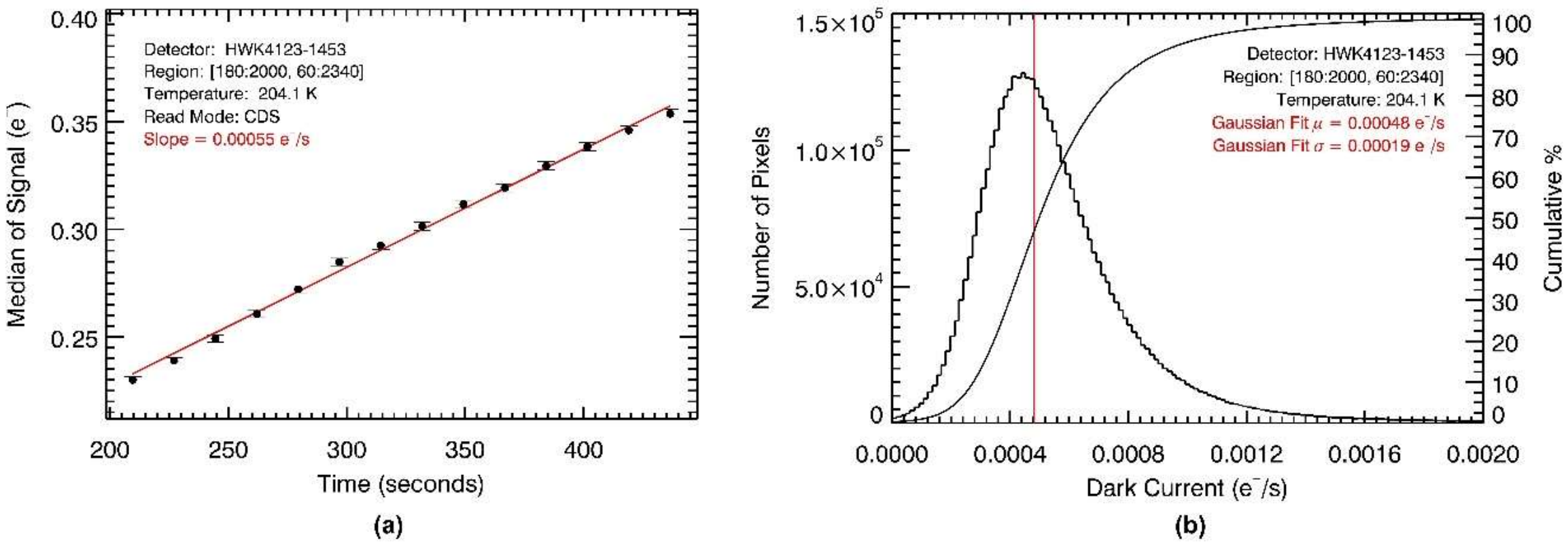

Figure 5. *(a)* The plot shows the median dark signal as a function of time at 204.1 K. The black points are the median dark signal of all pixels in the analysis region. The red line is a fit to the signal as a function of time. The slope of the red line is the DC of 0.00055 $e^-$/s/pixel. The error bars are the standard error on the median of the pixel population. *(b)* The plot shows the histogram of DC per pixel at 204.1 K. The y-axis on the left is the number of pixels per DC bin on the x-axis. The y-axis on the right is the cumulative percentage of pixels up to a DC bin. The vertical red line is the mean of a Gaussian fit to the distribution, with a DC value of 0.00048 $e^-$/s/pixel.

### *3.2. Read Noise and Bit Error Rate*

The total noise and bit error rate distributions are shown in the left and right panels of Figure 6, respectively. Both panels show histograms overlaid with a cumulative percentage curve and the mean of a Gaussian fit (red).

The data consist of 10,000 dark frames at 32× gain, with an integration time of 346 μs at 2.5 fps. The total noise distribution is dominated by RN because DC shot noise is negligible at 346 μs. The mean and standard deviation of the Gaussian fit to the histogram yield an RN of $0.19 \pm 0.02$ $e^-$/pixel. Approximately 90% of the pixels have an RN below 0.3 $e^-$/pixel and satisfy the photon-counting threshold. The distribution has a hot tail characteristic of CMOS SPCDs [5] [6]. As the programmable gain decreases, the RN increases to $0.22 \pm 0.02$ $e^-$ at 16× gain, $0.32 \pm 0.03$ $e^-$ at 8× gain, and $2.13 \pm 0.12$ $e^-$ at 1× gain. The HWK4123 read noise satisfies the high-sensitivity UV/VIS instrumentation specification (<2.5 $e^-$) at all programmable gain settings, but is twice the visible coronagraph channel RN specification (<0.1 $e^-$) at 32× gain [3].

We derive the BER distribution by applying Equation 34 of [24] to the total noise distribution. For the data presented in Figure 6, 63% of pixels have a BER below 1%, meaning fewer than 1 in 100 measurements from those pixels are incorrectly reported. As explained in Section 4, the BER can be used to map floating-point measurements to their corresponding integer electron counts. Following this rounding step, the effective measurement uncertainty is governed by the variance of a Bernoulli trial rather than the analog read noise.

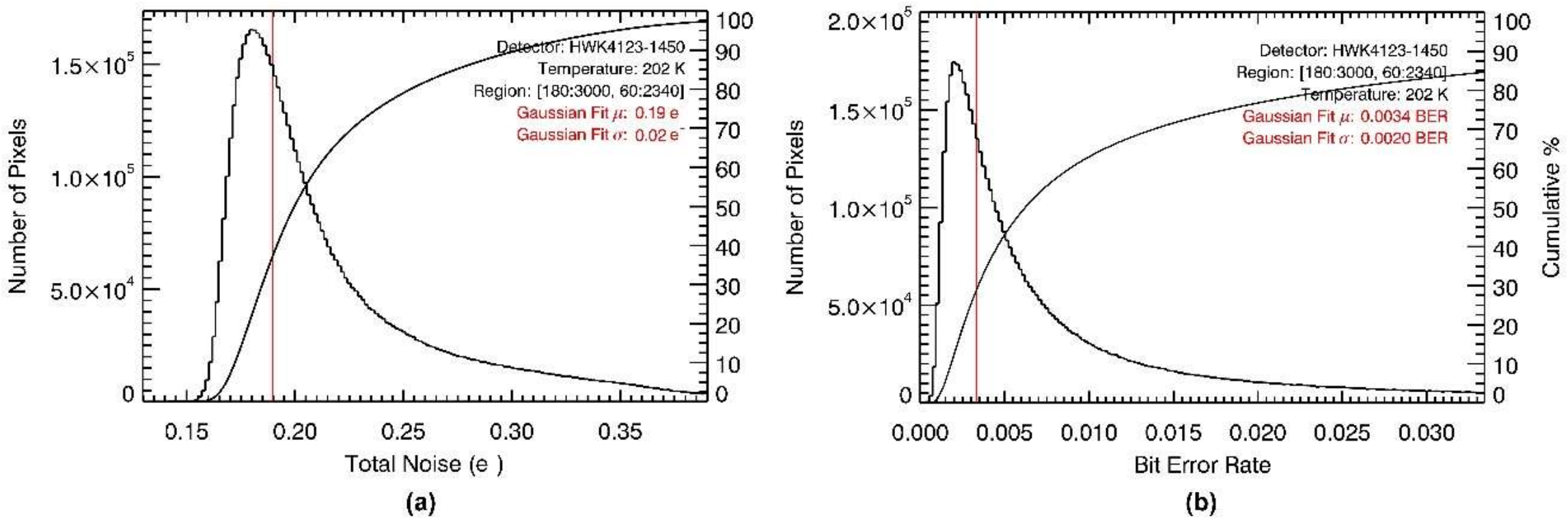

Figure 6. Both panels show a histogram where the left y-axis gives the number of pixels per bin, the right y-axis gives the cumulative percentage of pixels, and the red line marks the mean of a Gaussian fit. *(a)* The plot shows the total noise histogram. Since DC shot noise is negligible at 346 μs, RN dominates the total noise. As such, the Gaussian fit gives an RN of 0.19 $e^{-}$/pixel. *(b)* The plot shows the bit error rate based on the total noise histogram. The Gaussian fit gives a BER of 0.34%, and 63% of pixels have a BER below 1%.

### *3.3. Conversion Gain and FD Capacitance*

We measure $CG_{sys}$ using the PTE to get the inverse slope of a fit to the linear region of the PTC. Figure 7 shows PTCs for all programmable gains of the HWK4123. The black curve (dots) is 32× gain, the dark blue curve (squares) is 16× gain, the light blue curve (triangles) is 8× gain, and the red curve (crosses) is 1× gain. Each dot in the curves represents the median dark-subtracted signal for all pixels in the analysis region. The PTCs show that the detector output value reaches the maximum of the ADC sampling range (4,095 ADU), which is limited by the voltage swing at the FD and does not reach physical saturation [35]. Note that the x-axis in Figure 7 shows the dark-subtracted signals. We divide 4,095 ADU by $CG_{sys}$ to obtain the number of electrons at the maximum output value, listed as $N_{elec}$ in Table 4. Comparing the $N_{elec}$ and read noise values as a function of programmable gain shows that the HWK4123 trades a smaller well capacity for lower read noise. The team is investigating methods to measure physical saturation at all gains by changing the voltage level at the TG.

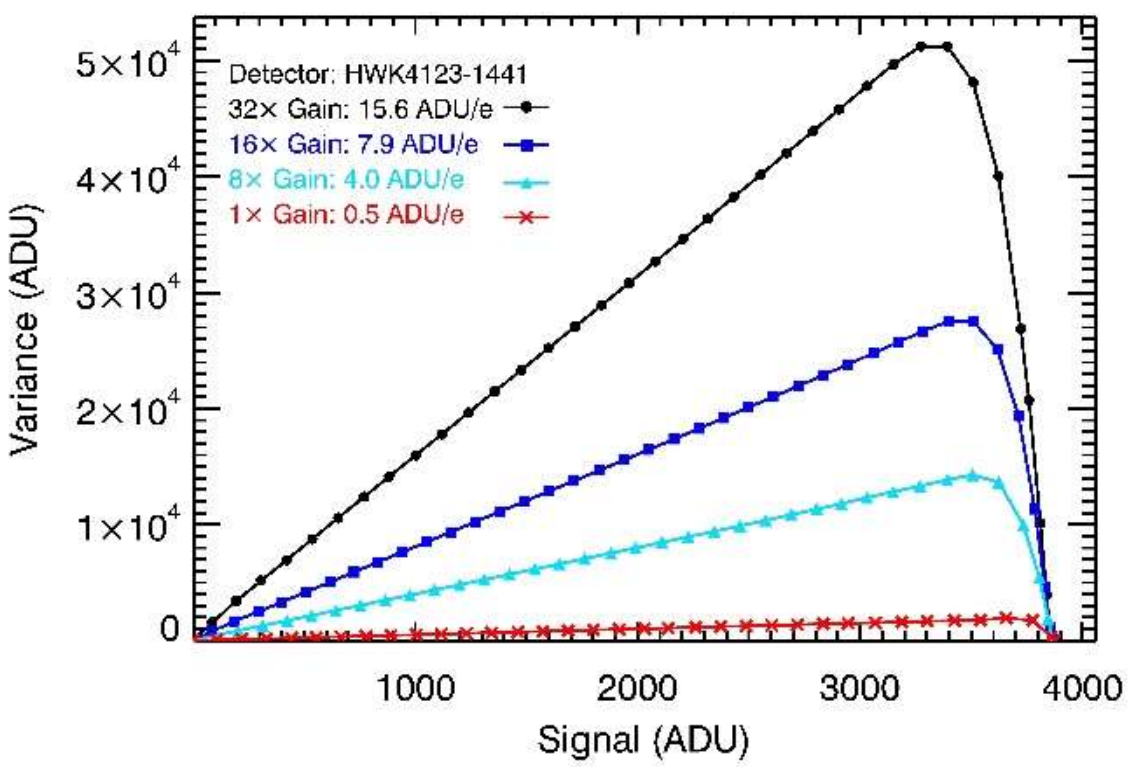

Figure 7. The plot shows PTCs for all programmable gains. The y-axis is the variance in signal in ADU. The x-axis is the signal in ADU. The inverse slope of the linear region is $CG_{sys}$.

We measure $CG_{measured}$ and multiply by $CG_{sys}$ to calculate $CG_{FD}$, using Eq. 6 and the inferred value of $\varepsilon_{coupling} = 0.94$ (see Section 2.1.3 for derivation using the manufacturer's $CG_{FD}$). We derive $CG_{elec}$ using Eq. 5, $CG_{measured}$, $\varepsilon_{coupling} = 0.94$, and $G_{SF} = 0.9$. We calculate $C_{FD}$ using Eq. 7 and $CG_{FD}$. Table 4 compiles measurements of $CG_{sys}$ and $CG_{measured}$, and derived values for $CG_{elec}$, $CG_{FD}$, $C_{FD}$, and $N_{elec}$ for all programmable gains of the HWK4123.

|  | $CG_{sys}$ [ADU/e⁻] | $CG_{measured}$ [μV/ADU] | $CG_{elec}$ [μV/ADU] | $CG_{measured}$ ×$CG_{sys}$[μV/e⁻] | $CG_{FD}$ [μV/e⁻] | $C_{FD}$ [fF] | $N_{elec}$ [e⁻] |
|---|---|---|---|---|---|---|---|
| Gain: 32× | 15.6 | 11.6 | 9.8 | 181.2 | 170.3 | 0.94 | ~256 |
| Gain: 16× | 7.9 | 22.8 | 19.3 | 180.5 | 169.7 | 0.94 | ~504 |
| Gain: 8× | 4.0 | 45.6 | 38.6 | 180.6 | 169.8 | 0.94 | ~1,010 |
| Gain: 1× | 0.5 | 361.1 | 305.5 | 180.6 | 169.8 | 0.94 | ~8,000 |

Table 4. This table lists measurements for $CG_{sys}$ and $CG_{measured}$, and derived values for $CG_{measured}$ × $CG_{sys}$, $CG_{elec}$, $CG_{FD}$, $C_{FD}$, and $N_{elec}$, for all programmable gains of the HWK4123.

### *3.4. Linearity*

We report the low-signal nonlinearity (LSNL) in the HWK4123 with the PHOTRON and QUEST systems. Using the PTE, we measure the detector response as a function of integration time under constant illumination (Figure 8) and determine the deviation from the ideal linear response for different gain settings (Figure 8 and Figure 9). Previous studies have reported nonlinearity for the QUEST camera operating in dual-gain mode [38] [42] [20]. Here, we characterize the HWK4123 with PHOTRON at 32× and 1× gain, and with QUEST in dual-gain mode.

For the QUEST system, we identify the linear regime between 2,400 ADU and 62,000 ADU. The lower bound is set by the discontinuity associated with the transition from high to low gain as signal increases, while the upper bound avoids saturation, which occurs at 65,335 ADU. These bounds are similar to those used in [38]. For PHOTRON, we identify the linear regime from 1,000 to 3,000 ADU, which represents 25% to 75% of saturation (4,095 ADU). For each system, we use the PTC to compute the system conversion gain in the linear regime to convert the measured signal from ADU to e⁻. We find 7.88 ADU/e⁻ for QUEST, 15.56 ADU/e⁻ for PHOTRON at 32× gain, and 0.49 ADU/e⁻ for PHOTRON at 1× gain.

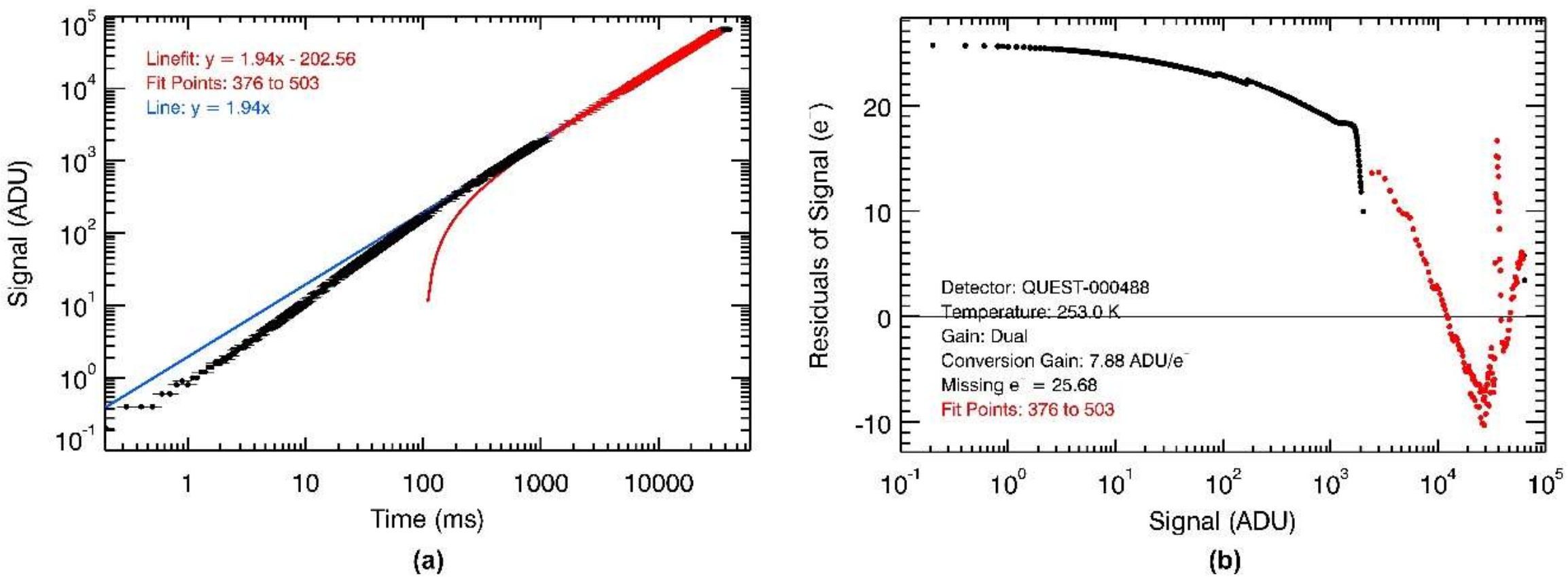

Figure 8. *(a)* The plot shows data from the PTE taken with QUEST. The black and red dots are the signal in ADU. The red dots are in the linear regime of the QUEST, to which we fit a line. The slope of this line represents the ideal linear response, shown in blue. The error bars are the standard error on the median value of all pixels per frame. *(b)* The black and red dots are the residuals of subtracting the measured signal from the ideal linear response, in electrons. The red dots are in the linear regime. The residuals suggest a maximum deviation of 26 e⁻ at low signals.

We calculate a maximum deviation of 26 e⁻ for QUEST, in agreement with the 30 e⁻ reported by [38]. For PHOTRON, the maximum deviation depends on the programmable gain, reaching 5 e⁻ at 32× gain and 71 e⁻ at 1× gain (Figure 9). The signal deviation does not scale linearly with programmable gain, which would be expected if LSNL originated entirely before the programmable gain stage, where gain changes would scale the LSNL uniformly. Consequently, the LSNL gain dependence suggests that the mechanism responsible for LSNL is at least partially associated with processes after the programmable gain stage. Therefore, charge-trapping or incomplete charge transfer may not fully explain the observed LSNL. However, [38] found that increasing the detector temperature reduced the impact of LSNL. Trap lifetimes decrease with higher temperatures, suggesting that traps may be present and contribute to LSNL. Our team will investigate the effects of temperature on LSNL for QUEST and PHOTRON.

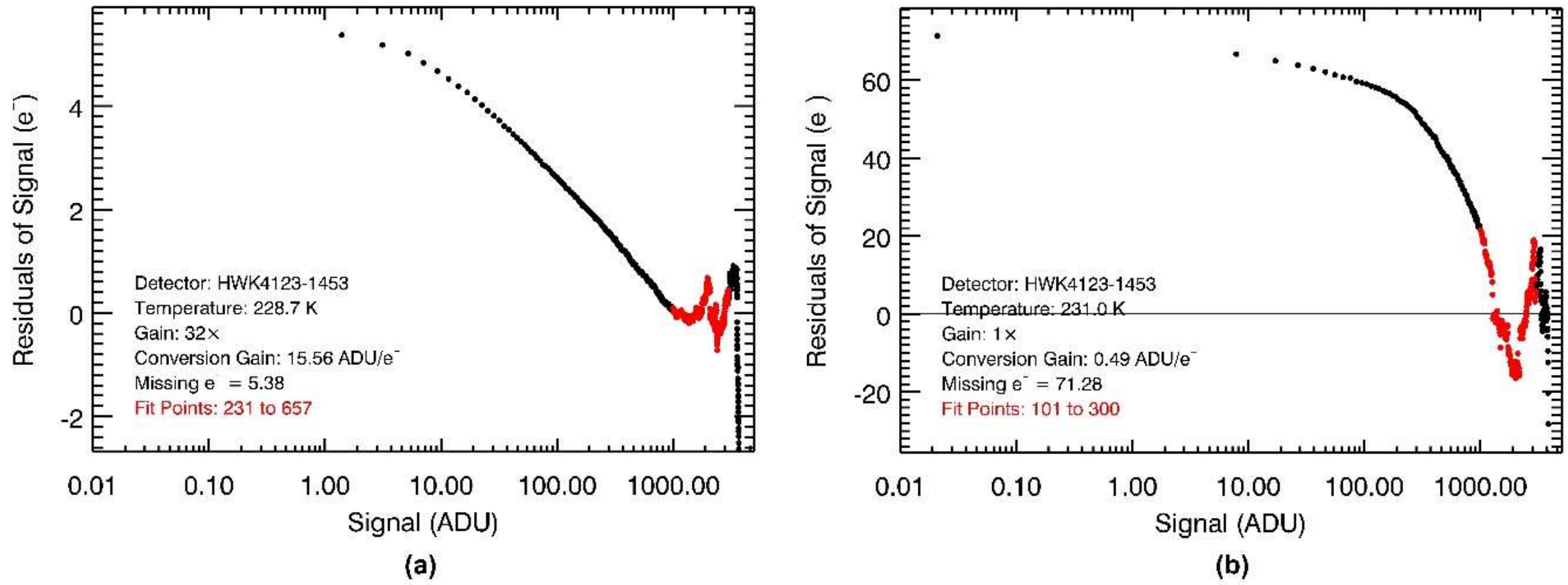

Figure 9. Both panels show results for the residuals of subtracting PHOTRON PTE data from the ideal detector response derived from each PTE. The red dots are data for which the PTE is linear. *(a)* The residuals for data taken at 32× gain suggest a maximum deviation of 5 $e^-$ at low signals. *(b)* The residuals for data taken at 1× gain suggest a maximum deviation of 71 $e^-$ at low signals.

### 3.5. *Quantum Efficiency*

The QE shown in Figure 10 is measured with the QUEST system, which is in agreement with [15]. The peak QE is 88% at 485 nm as shown by the red line. The central wavelengths of the measurements have a spacing of 5 nm. Each black dot in Figure 10 represents the median QE for all pixels in the analysis region, for an average of 10 dark-subtracted frames at 253 K. At wavelengths shorter than 400 nm, the probability of producing more than one electron increases (QY greater than unity). The team is actively investigating QY corrections to QE below 400 nm by following the work of [39]. The results shown here are not corrected for linearity. However, the data are taken in the linear regime of the QUEST, and should not shift more than 5%. Similarly, data taken with PHOTRON above 450 nm are in the linear regime and agree with published values and with Figure 10, for the temperatures listed in Table 3 [35]. For PHOTRON, data below 450 nm were not taken in the linear regime and require a linearity correction, as the measured QE is lower than published values. The HWK4123 QE partially satisfies the specifications for the high-sensitivity UV/VIS instrumentation (> 80%) and for the visible coronagraph channel (> 50%) [3].

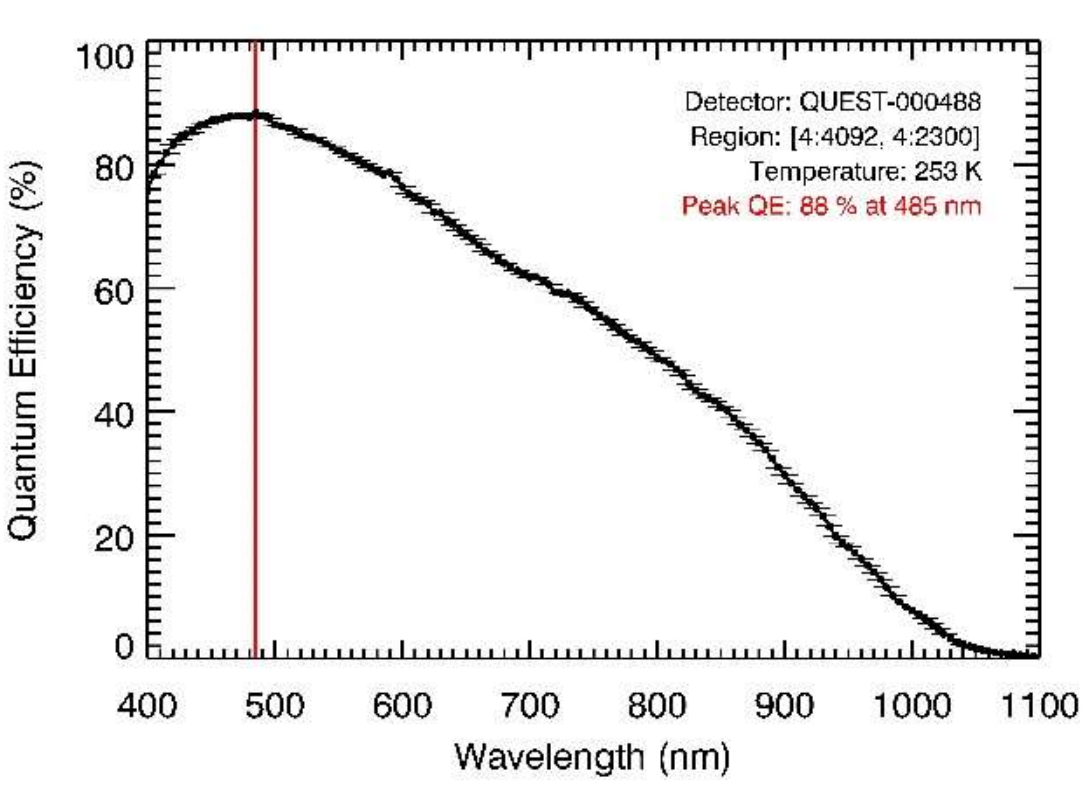

Figure 10. The curve shows the HWK4123 QE measured at 253 K with the QUEST system. The red line marks the peak QE of 88% at 485 nm. The error bars are the standard error on the median of the pixel population.

Using the QE experiment, we also obtain preliminary measurements of the sensor flat-field uniformity. We acquired and averaged 25 frames at a wavelength of 800 nm with PHOTRON operating at 32× and 1× gain (Figure 11). The mean QE is denoted with μ, and the spatial standard deviation with σ. The data are in the linear regime of the sensor, but are not corrected for linearity. Consequently, the results reported in Figure 11 are tentative and motivate further study alongside the LSNL characterization. These tentative results suggest that the sensor may have slightly higher QE and flat-field uniformity at 1× gain.

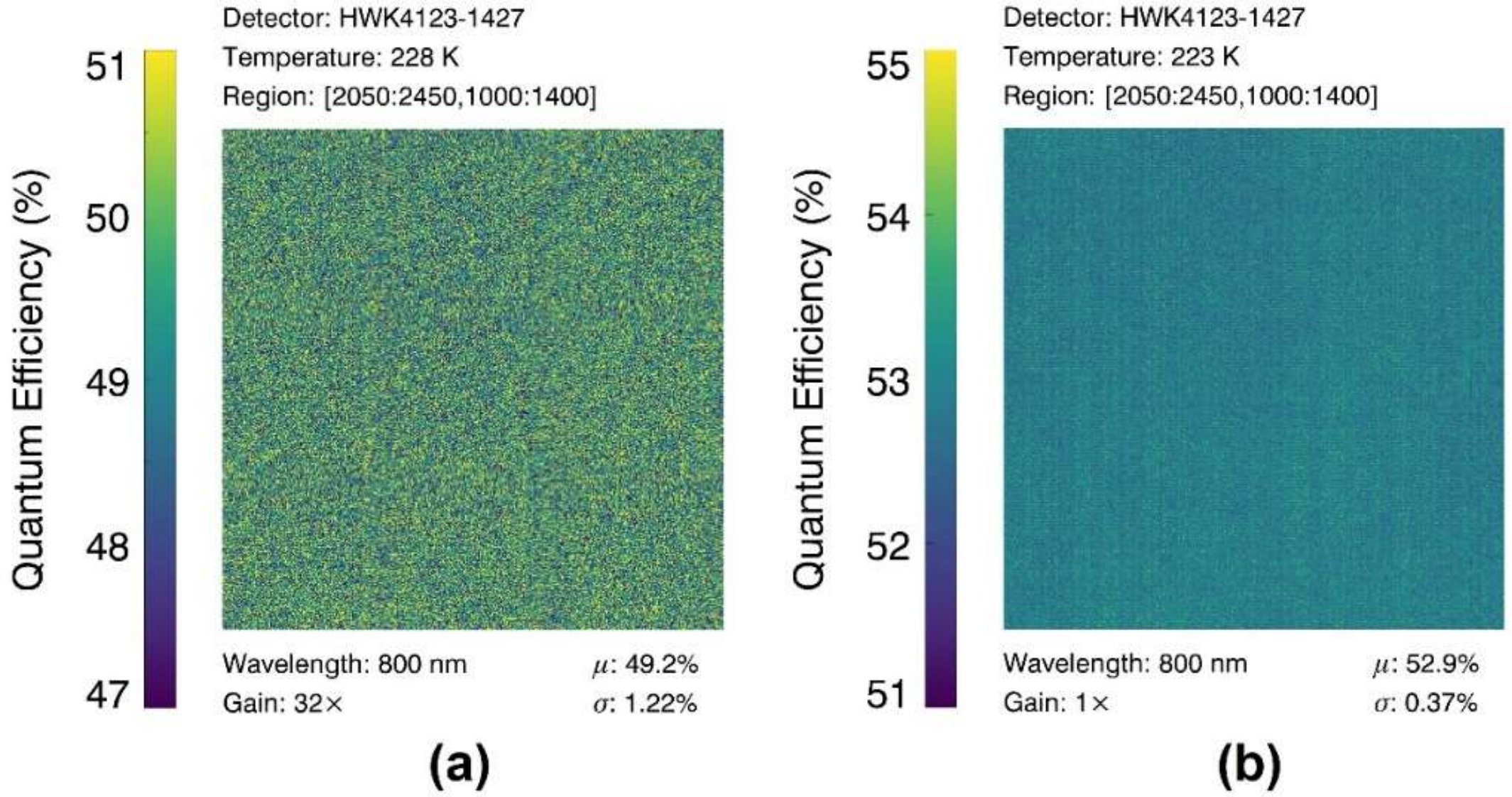

Figure 11. Both panels show the HWK4123 QE map at 800 nm, using PHOTRON, for operation at *(a)* 32× gain and *(b)* 1× gain. The mean QE is labeled as μ, and the spatial standard deviation as σ. The data are not corrected for linearity, but are taken in the linear regime of the detector. As such, the differences in QE and flat field uniformity as a function of programmable gain are tentative and motivate further study.

### *3.6. Crosstalk*

We measure crosstalk using the proportion of the signal in pixels surrounding a central pixel affected by an HEE. Figure 12 shows the mean crosstalk values for ~1($10^5$) events. Crosstalk is negligible for the HWK4123 when compared to dominant noise terms under assumed observing conditions (*i.e.*, dark current, read noise, zodiacal and exozodiacal dust) and will not be a concern for a flagship mission such as HWO.

Detector Temperature: 235 K
Number of Images: 100
Number of Events: 103531
Region: [200:4000,100:2300]
Crosstalk Results (%):

| | | |
|---|---|---|
| 0.00 | 0.23 | 0.00 |
| 0.20 | 99.15 | 0.16 |
| 0.00 | 0.26 | 0.00 |

Figure 12. The panel shows the crosstalk for the HWK4123.

### *3.7. Persistence*

We measure persistence as the residual signal in a pixel following a reset after exposure. We use a pulsed laser mounted on the QUEST camera to illuminate the HWK4123 to an initial fluence of ~5100 $e^-$ (~70% of saturation) in the central 500×500 pixels. Each pulse set comprises ten 0.5 s frames, the first of which contains the 50 ms pulse, and the latter nine are dark. We subtract dark current and bias from all frames. We acquired 49 pulse sets and averaged the data for each corresponding frame in the sequence. Only 0.026% of the initial signal remains (~1 $e^-$) in the first dark frame after the pulse (0.5 s), while the residual signal in the second dark frame after the pulse (1 s) decreases to 0.002% (~0.1 e⁻), below the detector read noise. Figure 13 shows the histograms for the first and second dark frames after the pulse. This level of persistence is negligible, particularly since target slews in space observatories (such as HWO) provide ample time for repeated detector readout and reset between observations. For a given observation, the impact of persistence on SNR depends on target brightness and integration time.

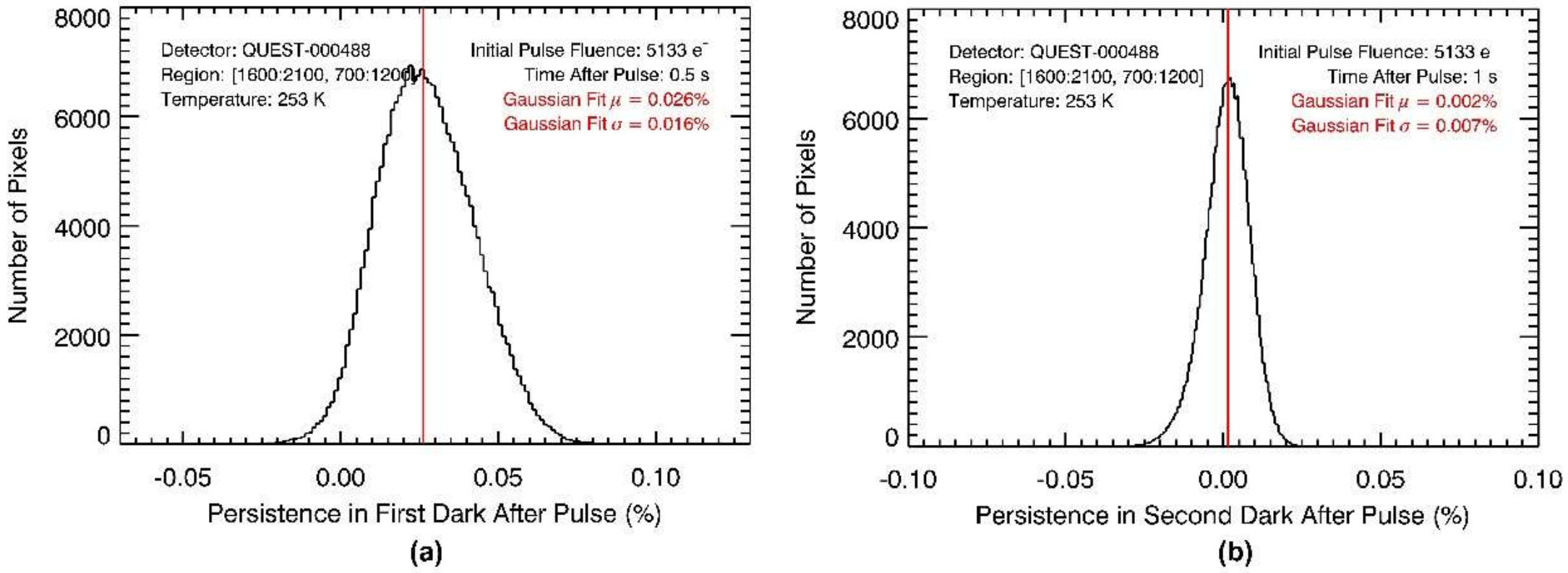

Figure 13. Both panels show a histogram of persistence in the pixel population after exposure to a pulsed laser. The red line is the mean of a Gaussian fit. *(a)* In the first dark frame after the pulse, only 0.026% (~1 e⁻) of the initial ~5100 e⁻ remain. *(b)* In the second dark frame after the pulse, the 0.002% (~0.1 e⁻) remaining signal is below the read noise of the detector.

### *3.8. Glow*

We measure glow in a small population of a few hundred reference pixels, clustered in the detector corners. Figure 14 shows a representative sample of the glow population in the bottom left corner of the sensor. For this DC measurement at 184.5 K, the glowing reference pixels have a median DC of 0.0016 e⁻/s, four times higher than the median DC of the imaging pixels. The histograms compare a sample of imaging pixels (black) to the glowing reference pixels (red) contained in a 100×5 pixel region in the corner. The imaging pixel population includes hot pixels that are scattered across the sensor, forming the hot tail in the histogram. In contrast, the glowing reference pixels are clustered in the corners, suggesting multiplexer glow from surrounding circuitry. These pixels do not affect the active imaging region used in analysis, and do not affect reported performance metrics. We do not identify additional spatial structure in the sensor array DC map.

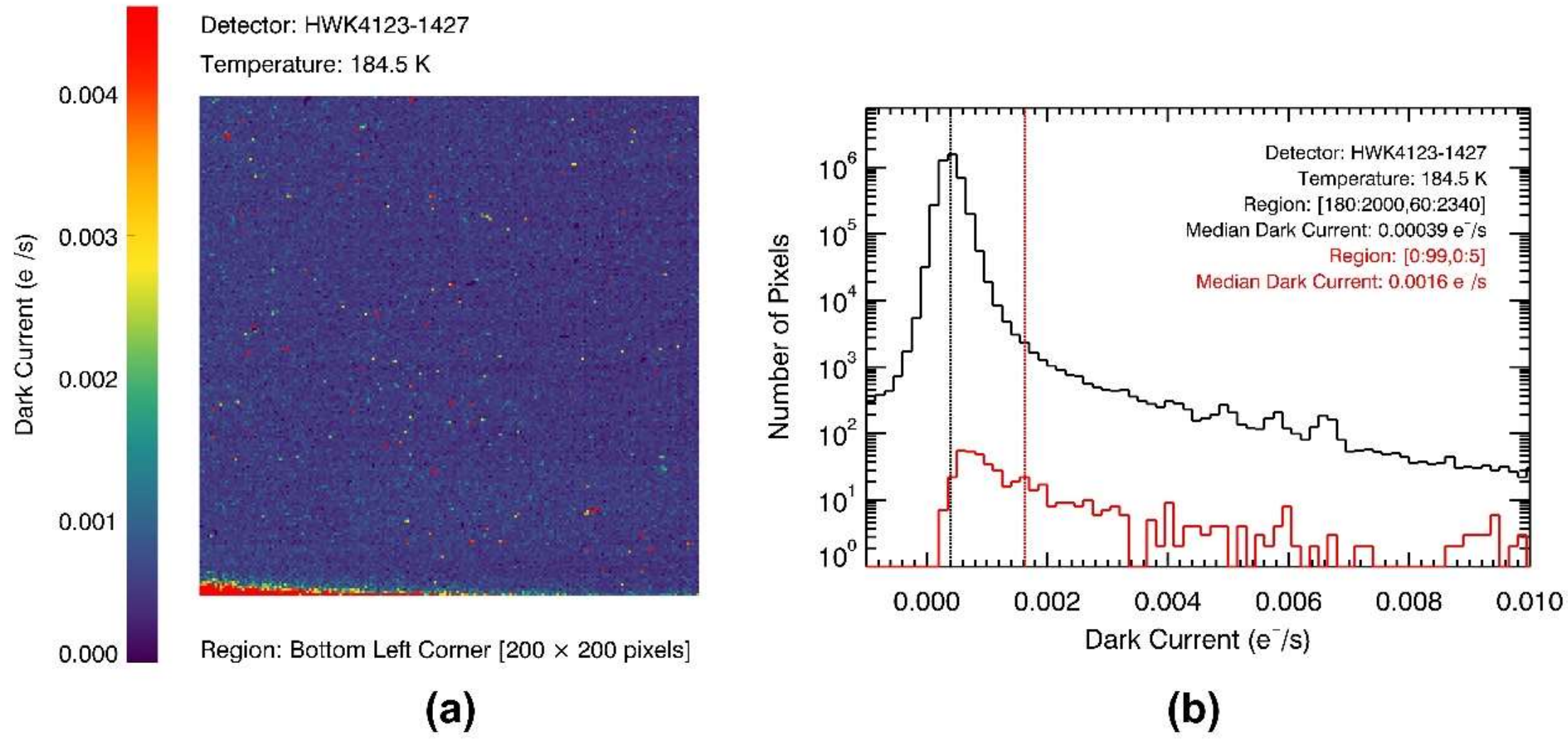

Figure 14. *(a)* The DC map shows a 200×200 corner of the HWK4123. There is noticeable glow in the 100×5 pixel region in the bottom left corner. *(b)* The histograms show the DC population for a sample of imaging pixels in black, and glowing reference pixels in red. The vertical lines show the median values.

In contrast, a spatially uniform unit-cell glow component would contribute an effective dark current without leaving a distinct spatial signature. If this component varies weakly with temperature, it may contribute to an apparent dark current plateau at lower temperatures [43] [44]. One way to measure unit-cell glow in infrared detectors is to take advantage of up-the-ramp (UTR) sampling, as described in [43]. Unfortunately, the HWK4123 does not offer UTR. We are investigating methods to characterize potential unit-cell glow for destructive CDS reads.

## 4. Simulated Low-Flux Observation

This section describes the impact of detector performance on the SNR of an idealized low-flux exoplanet observation. Following the work of [45] [46] [47], we estimate the integrated photon flux per spectral channel for modern Earth, shown in Figure 15. We use a 6-m diameter telescope and an end-to-end throughput of 27% [48]. On Earth, the presence of $O_2$ is due to a biological process: photosynthesis. While other life-bearing eons of Earth lacked detectable $O_2$, it reveals the presence of our modern biosphere and may do so for other exoplanets, with a careful treatment of false positives [46] [49] [50] [51] [52] [53] [54]. As such, we select the $O_2$-A band at 760 nm as the target biosignature, and obtain ~8 photons per hour from the line center.

We calculate SNR as a function of time for a single pixel. Since the HWO design is not finalized, we use the simulation above to justify a photon arrival rate of ~1 photon per hour per pixel to decouple the SNR calculation from unknowns. This analysis ignores contaminating sources of photons (zodiacal and exozodiacal dust, coronagraph speckles, scattered starlight) to focus on the impact of detector performance on SNR when observing at a rate of one photon per hour. Table 5 describes the quantities used to calculate SNR in Eq. 9.

We assume an integration time of 250 s per frame to limit the effects of cosmic rays to ~1% of the pixel array. HWO will orbit at the Sun-Earth Lagrange Point 2 (L2), where JWST measurements show a shielded cosmic ray event rate of ~2.3 ions/cm$^2$/s near solar maximum [55]. A typical event affects 7.1 pixels in JWST's 18 μm-pitch H2RG detectors, which scales by area to ~100 HWK4123 pixels. The HWK4123 has an active area of ~2 cm$^2$, implying ~1150 cosmic ray events in a 250 s exposure and contamination of ~1% of its 9.4 Mpixels. We calculate the number of frames (n) by dividing the total integration time (t) by 250 s, and use a step size of 250 s to avoid fractional frames.

| Symbol | Quantity Description | Units | Notes |
|---|---|---|---|
| $S_s$ | Signal from space | γ/s/channel | ~1 γ/hr/pixel |
| QE | Detector quantum efficiency | $e^-$/γ | from Section 3.4 |
| DC | Detector dark current | $e^-$/s/pixel | from Section 3.1 |
| RN | Detector read noise | $e^-$/pixel | from Section 3.2 |
| n | Number of frames | dimensionless | n = t / 250 s |
| t | Total integration time | s | step size = 250 s |

Table 5. This table lists the quantities used to determine SNR in Eq. 9.

$$SNR(t,n) = (S_s \cdot t \cdot QE)/\sqrt{(S_s \cdot QE + DC) \cdot t + RN^2 \cdot n} \quad \text{Eq. 9}$$

Figure 15 shows the SNR for a single pixel observing one photon per hour under multiple detector assumptions. The first two curves correspond to a perfect detector (solid light blue) and a detector with HWO's specifications from Table 6 (dashed dark blue). The perfect detector, with zero RN, zero DC, and 100% QE, reaches SNR = 5 in 1.04 days, while the HWO detector requires 1.81 days.

The remaining three curves correspond to the HWK4123 detector with its measured DC and QE performance, using three different treatments of RN: the measured RN performance (black dotted), zero RN (gold dashed), and a BER rounding step (red double dot). We use the BER rounding step to exploit an advantage of SPCDs by treating the reported digital value as the outcome of a Bernoulli trial.

As RN decreases, the BER drops rapidly (Figure 6), increasing the probability that the reported floating-point value lies closest to the correct integer number of electrons. Rounding to the nearest integer changes the variance of the signal estimator from analog read noise to the variance of a Bernoulli trial. This variance is p · (1 − p), where p is the probability of success, and (1 − p) is the probability of failure (*i.e.*, the BER). Consequently, we can rewrite the variance as p · (BER), and since 1 − p = BER, we substitute p with 1 − BER to get (1 − BER) · (BER). We can replace the $RN^2$ term in Eq. 9 with (1 − BER) · (BER), or simply BER as the first term is approximately 1 for small BER.

Using this treatment, the rounding-step HWK4123 case reaches SNR = 5 in 8.20 days, approaching the 8.09 days for the zero-RN HWK4123 case, and improving upon the 9.88 days required for the regular HWK4123 performance. The BER rounding step does not eliminate the detector's analog RN, but instead takes advantage of a concept similar to the binary symmetric channel in telecommunications to significantly reduce noise. Even so, current CMOS SPCDs require longer integration times than the HWO target detector, reducing the total number of observations possible for a fixed mission lifetime.

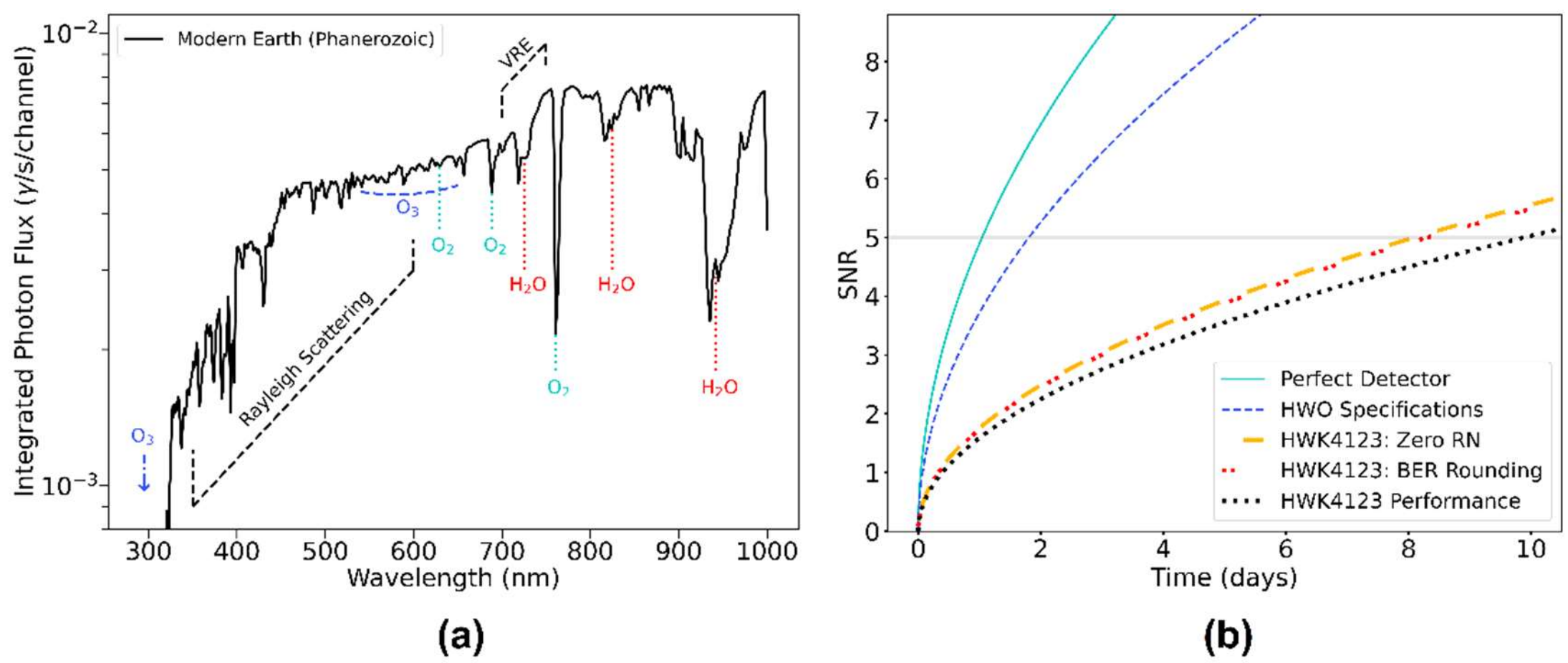

Figure 15. *(a)* The plot shows the integrated spectrum of modern Earth with the $O_3$, $O_2$, $H_2O$, and vegetation red edge (VRE) biosignatures. The y-axis is the number of photons per second per spectral channel on the x-axis. *(b)* The plot shows SNR as a function of time for different detector assumptions.

## 5. Discussion

The current HWO concept design has instruments that span a range of wavelengths [56]. Table 6 lists a selection of the most stringent detector performance specifications found in the HWO Roadmap [3]. The table is organized by three detector metrics and three spectral regions.

| | DC [e⁻/s/pixel] | RN [e⁻/pixel] | QE |
|---|---|---|---|
| NUV/visible | < 0.002 | < 2.5 | > 50% |
| Visible | < 0.0001 | < 0.1 | > 80% |
| NIR | < 0.001 | < 0.1 | > 90% |

Table 6. This table lists the most stringent performance specifications for HWO detectors [3].

The HWO Roadmap identifies prospective detector technologies for the categories in Table 6. The NUV/visible detector candidates are CMOS SPCDs, electron-multiplying charge-coupled devices (EMCCDs), skipper-CCDs, and microchannel plates (MCPs). However, MCPs have low quantum efficiency, such as GaN-b12 photocathodes with a QE of ~ 35 to 10% spanning the 200 to 400 nm range [57]. The visible range candidates are CMOS SPCDs, EMCCDs, and skipper-CCDs.

### *5.1. HWK4123 Performance and Other State-of-the-Art Detectors*

The following section compares the measured performance of the HWK4123 to the current HWO specifications at Technology Readiness Level 5, and contextualizes the HWK4123 with other state-of-the-art detectors in the NUV/visible and visible range, listed in Table 7.

The HWK4123 performance values partially satisfy HWO's current detector specifications. The DC and RN are well below the NUV/visible specifications, but are ~5× and ~2× larger than the threshold for the visible range coronagraph, respectively. The HWK4123 QE satisfies the visible range specifications from 410 to 570 nm. We provide suggestions to improve DC, RN, and QE in Section 6.

The HWK4123 performance is comparable to the QIS, EMCCDs, and skipper-CCDs for the values listed in Table 7. However, skipper-CCDs can read nondestructively and lower the read noise of a measurement with correlated multiple sampling. The pixel architecture of the HWK4123 does not permit nondestructive reads, but future CMOS SPCD designs can account for this.

| Detector | Pixel size [µm] | DC [e⁻/s/pixel] | RN [e⁻/pixel] | Peak QE |
|---|---|---|---|---|
| HWK4123 (CMOS SPCD) | 4.6 | 0.0005 at 204.1 K | 0.19 at 2.5 fps | 88 % (485 nm) |
| HWK1411 (CMOS SPCD) [13] | 8.0 | 2.3 at 293 K | 0.5 at 120 fps | 80 % (600 nm) [35] |
| QIS (CMOS SPCD) [5] | 1.1 | 0.0005 at 253 K | 0.27 | 83 % (500 nm) |
| QIS (CMOS SPCD) [12] | 1.1 | 1.12 at 333 K | 0.35 | 87 % (480 nm) |
| Deep-depletion EMCCD [58] | 13 | 0.0005 at 165 K | 0.2 | 92 % (550 nm) |
| Nüvü EMCCD [59] | 16 | 0.0001 at 183 K | 0.1 | 95 % (575 nm) |
| EMCCD (e2v CCD201-20) [60] | 13 | 260 at 293 K | < 1 | 95 % (575 nm) |
| EMCCD (e2V CCD97) [61] | 16 | 0.000072 at 188 K | < 1 [62] | 96 % (575 nm) [62] |
| SENSEI skipper-CCD [63] | 15 | $6.8\times10^{-9}$ at 135 K[a] | $2.5/\sqrt{n_{sample}}$ | N/A |
| AstroSkipper skipper-CCD [64] | 15 | 0.0002 at 140 K | $4.3/\sqrt{n_{sample}}$ | 99 % (880 nm) |
| LBNL skipper-CCD [65] | 15 | $2.3\times10^{-9}$ at 140 K[b] | $3.57/\sqrt{n_{sample}}$ | N/A |
| Skipper-CCD [66] | 10 | 0.00028 at 143 K | 0.2 | N/A |

Table 7. This table summarizes the performance of state-of-the-art HWO detector candidates.
[a] value calculated from [63]
[b] value calculated from [65]

It is important to consider other detector performance metrics beyond those compiled in Table 7, such as the detector operating conditions, the pixel and readout architecture, and noise specific to a technology. CMOS SPCDs offer significant advantages over EMCCDs and other photon-counting technologies for space instruments. CMOS SPCDs have deep sub-electron read noise through small-capacitance sense nodes and direct-readout architectures [67] [68]. This enables single-photon detection and photon-number resolution at room temperature without requiring avalanche gain [69].

Unlike CCD technologies, CMOS devices use a direct-readout architecture that eliminates pixel-to-pixel charge transfer. This avoids radiation-induced charge transfer inefficiency, which is critical for long-duration NASA missions in high-radiation environments, such as L2 or the Jovian system [68] [70] [71]. In CCDs, radiation-induced traps capture and release signal charge during transfer, causing image trailing, flux loss, and systematic errors, particularly for faint objects. For example, radiation-induced charge traps within the Hubble Space Telescope’s Advanced Camera for Surveys Wide Field Channel capture and release signal charge during pixel-to-pixel transfer [72] [73] [74]. This disproportionately affects the pixels farthest from the readout amplifier, which need more transfers to read out charge and lose more than 50% of their electrons under faint observing conditions. CMOS SPCDs avoid these effects throughout the mission lifetime without requiring complex correction algorithms.

Compared to EMCCDs, CMOS SPCDs provide single-photon sensitivity with higher radiation tolerance and simpler operation. EMCCDs amplify signal electrons via a high-voltage gain register, which is vulnerable to radiation damage and single-event transients. EMCCDs also experience gain aging, clock-induced charge, and radiation-induced charge transfer inefficiency, all of which reduce quantitative photon counting over time [68] [70] [71]. CMOS SPCDs achieve photon sensitivity through high conversion gain at the pixel level, preserving Poisson statistics, supporting high frame rates, and operating near room temperature without excess noise or long-term degradation.

EMCCDs and skipper CCDs have clock-induced charge (CIC) and charge transfer inefficiency, rely on serial charge transfer and off-pixel amplification, and consume more power to operate. In EMCCDs, CIC is amplified by the electron-multiplying stage. The CIC reported for the EMCCDs and skipper-CCDs in Table 7 ranges from 0.003 to 0.0009 $e^{-}$/pixel/frame [58] [59] [63] [64].

Relative to single-photon avalanche diodes, microwave kinetic inductance detectors (MKIDs), transition-edge sensors (TESs), and superconducting nanowire detectors (SNSPDs), CMOS SPCDs balance performance with system-level feasibility. Superconducting and avalanche-based detectors can achieve exceptional timing resolution and low dark counts but require cryogenic systems, specialized electronics, and smaller pixel formats. These requirements increase instrument cost, mass, and integration complexity, particularly for wide-field imaging.

CMOS SPCDs stand apart not only for photon-counting sensitivity and radiation resilience, but also for their system-level integration. Unlike other detector technologies that require extensive off-chip electronics, cryogenic cooling, or complex biasing, CMOS SPCDs incorporate the entire readout ecosystem on-chip, including power regulation, clock generation, programmable readout modes, and digitization. This system-on-a-chip approach reduces instrument size, weight, and power, while increasing reliability and flexibility. In practice, the detector and readout function as a single, self-contained unit that is easier to replace or upgrade in space, similar to the Hubble Space Telescope's SIDECAR controller, rather than relying on bulk electronics distributed across the spacecraft. By combining monolithic integration, scalability to gigapixel formats, near-room-temperature operation, and full single-photon sensitivity, CMOS SPCDs enable NASA instruments to detect the faintest sources with unprecedented efficiency and operational simplicity.

## 6. Roadmap to Improve CMOS SPCDs for HWO

The HWO Roadmap specifies the need to improve detectors to enable scientific applications. The following sections provide pathways to improve CMOS SPCD read noise, dark current, and quantum efficiency.

Future CMOS SPCD designs can lower RN through targeted circuit modifications. For example, adopting buried channel output FETs could directly reduce noise contributions at the output stage [75] [76]. In addition, designers can implement readout modes that use a larger number of samples per pixel (such as correlated multiple sampling), thereby lowering the effective RN through averaging [77] [78].

Dark current reduction should focus first on identifying and mitigating glow contributions in the current device generation. If glow dominates, design changes can suppress it by adding localized shielding or blocking layers around the responsible circuit elements [79]. Designers could also consider transitioning to a hole-collecting pixel architecture, although past evidence is relatively sparse that such a choice has yielded lower intrinsic dark current. Additional reductions may be achievable through improved surface and trench passivation, applied consistently at the front side, back side, and etched interfaces.

One method to improve QE at longer wavelengths is to make thicker pixels to match the absorption length (for silicon-based devices, [80]). For instance, [35] compared the QE of the HWK4123 to the HWK1411 (also from Fairchild Imaging), which have a pixel thickness of 2.5 μm and 6.0 μm, respectively. The HWK1411 QE exceeds that of the HWK4123 by approximately ten percentage points from 600 to 1000 nm.

One technique to enhance QE at shorter wavelengths is backside engineering to modify the electric field in the silicon substrate and push charges away from surface traps towards the collection area. Delta-doping is such a technique and uses molecular beam epitaxy to dope the silicon layer with boron [81], with reported improvements for CCDs [82], EMCCDs [81], and CMOS sensors [83].

## 7. Conclusion

The Habitable Worlds Observatory will need improved detector technology to achieve its science goals. CMOS SPCDs satisfy some of the detector specifications for read noise, dark current, and quantum efficiency. Specifically, the HWK4123 is a photon-number-resolving detector with a dark current of 0.0005 $e^-$/s/pixel, a read noise of 0.19 $e^-$/pixel, a quantum efficiency above 80%, and negligible persistence and crosstalk. These results indicate that CMOS SPCDs represent a credible and technologically mature pathway toward development for HWO. Part of that development includes the completion of our ongoing on-sky testing program at a telescope, radiation testing for an equivalent irradiation dose at L2, where HWO would orbit, and the complete characterization of LSNL sources such that future CMOS SPCD designs can mitigate them. Proven methods exist to improve the DC, RN, and QE performance of silicon-based devices such as the HWK4123. Consequently, the continued development of CMOS SPCDs offers a clear path to reducing detector risk for future HWO instruments.

## Disclosures

The authors declare that there are no financial interests, commercial affiliations, or other potential conflicts of interest that could have influenced the objectivity of this research or the writing of this paper.

## Code and Data Availability

Data reported in this paper will be free for reuse. The Flexible Image Transport System (FITS) data contain significant intellectual property and will only be shared in response to requests that do not jeopardize confidentiality.

## Acknowledgments

The results presented in this paper are supported by NASA’s Strategic Astrophysics Technology (SAT) Program (Grant No. 80NSSC20K0310). The large language model ChatGPT was used for language and grammar clean-up.

## References

[1] "Pathways to Discovery in Astronomy and Astrophysics for the 2020s," National Academies Press, Washington, 2021.

[2] M. R. Bolcar, F. Zhao, P. Scowen and e. al., "Habitable Worlds Observatory Technology Development Plan," 29 July 2025. [Online]. Available: https://ntrs.nasa.gov/api/citations/20250007159/downloads/Bolcar_HWOTechRoadmaps_HWO25_20250729_v3.pdf?attachment=true.

[3] M. R. Bolcar, F. Zhao and P. Scowen, "Habitable Worlds Observatory Technology Roadmap," 14 11 2024. [Online]. Available:

https://ntrs.nasa.gov/api/citations/20240014305/downloads/Bolcar_MirrorTechDays_HWOTech_20241118_v3.pdf.

[4] J. P. Gallagher, L. Buntic, W. Deng, E. R. Fossum and D. F. Figer, "Radiation tolerance of a single-photon counting complementary metal-oxide semiconductor image sensor," *JATIS,* 2024.

[5] J. Gallagher, L. Buntic, E. Alexani, K. Bouthsarath and D. Figer, "Characterizing radiation-tolerant single photon resolving CMOS detectors," *Proceedings of the SPIE,* vol. 13103, 2024.

[6] J. P. Gallagher, L. Buntic, W. Deng and D. F. Figer, "Characterization of single-photon sensing and photon-number resolving CMOS image sensors," *SPIE Astronomical Telescopes + Instrumentation,* vol. 12191, 2022.

[7] J. P. Gallagher, "Characterization of Single Photon Sensing and Photon Number Resolving CMOS Detectors for Astrophysics," ProQuest LLC., Rochester, 2020.

[8] N. R. Shade, G. Kyne, S. Nikzad, E. Charbon and E. R. Fossum, "Characterization and validation of next generation image sensors for space applications," in *SPIE Astronomical Telescopes + Instrumentation*, Yokohama, Japan, 2024.

[9] Fairchild Imaging, "HWK4123 Product Brief," August 2025. [Online]. Available: https://fairchildimaging.com/hubfs/Product%20Briefs/MAN%200244_HWK4123_Rev%20A.pdf.

[10] J. Ma, D. Zhang, O. Elgendy and S. Masoodian, "A Photon-Counting 4Mpixel Stacked BSI Quanta Image Sensor with 0.3e- Read Noise and 100dB Single-Exposure Dynamic Range," in *2021 Symposium on VLSI Circuits*, Kyoto, 2021.

[11] J. Ma, S. Chan and E. R. Fossum, "Review of Quanta Image Sensors for Ultralow-Light Imaging," *IEEE Transactions on Electron Devices,* vol. 69, no. 6, pp. 2824-2839, 2022.

[12] J. Ma, D. Zhang, D. Robledo, L. Anzagira and S. Masoodian, "Ultra-high-resolution quanta image sensor with reliable photon-number-resolving and high dynamic range capabilities," *Scientific Reports,* vol. 12, 2022.

[13] Fairchild Imaging Systems, "HWK1411_NOV2024_VF.pdf," 25 Novemeber 2024. [Online]. Available: https://fairchildimaging.com/hubfs/Product%20Briefs/HWK1411_NOV2024_VF.pdf.

[14] Tucsen Photonics, "Aries 6504 Pro," Tucsen Photonics, 02 02 2026. [Online]. Available: https://www.tucsen.com/uploads/Aries-6504-Pro-Specifications-20260202.pdf.

[15] Hamamatsu Photonics K.K., "ORCA®-Quest qCMOS® camera C15550-20UP Technical note," May 2022. [Online]. Available: https://camera.hamamatsu.com/content/dam/hamamatsu-photonics/sites/documents/99_SALES_LIBRARY/sys/SCAS0154E_C15550-20UP_tec.pdf.

[16] Hamamatsu Photonics K.K., "ORCA-Quest 2 qCMOS camera C15550-22UP," June 2025. [Online]. Available: https://www.hamamatsu.com/content/dam/hamamatsu-photonics/sites/documents/99_SALES_LIBRARY/sys/SCAS0166E_C15550-22UP.pdf.

[17] Hamamatsu Photonics K.K., "ORCA-Quest IQ qCMOS camera C15550-23UP," August 2025. [Online]. Available: https://camera.hamamatsu.com/content/dam/hamamatsu-photonics/sites/documents/99_SALES_LIBRARY/sys/SCAS0179E_C15550-23UP.pdf.

[18] C. Layden, K. Burdge, G. Furesz, J. Garcia-Mejia, J. Dinsmore, G. Mo, D. Osip, J. J. Piotrowski, R. W. Romani, A. Berne, D. Chakrabarty and E. Chickles, "proto-Lightspeed: a high-speed, ultra-low read noise imager on the Magellan Clay Telescope," *arXiv e-prints,* 2026.

[19] A. Roy, S. Feldman, P. Klupar, J. DiPalma, S. Perlmutter, E. S. Douglas, G. Aldering, G. Furesz, P. Ingraham, G. Stefansson, D. Kelly, F. Y. Yang, T. Wevers, N. Arulanantham, J. Lasker and M. Rigault, "The Lazuli Space Observatory: Architecture & Capabilities," *arXiv e-prints,* 2026.

[20] M. Lucas, B. Norris, O. Guyon, M. Bottom, V. Deo, S. Vievard, J. Lozi, K. Ahn, J. Ashcraft, T. Currie, D. Doelman, T. Kudo, L. Leboulleux and L. Lillley, "Visible-light High-contrast Imaging and Polarimetry with SCExAO/VAMPIRES," *Publications of the Astronomical Society of the Pacific,* vol. 136, no. 11, 2024.

[21] W. Deng, D. Starkey, J. Ma and E. R. Fossum, "Modelling Measured 1/f Noise in Quanta Image Sensors (QIS)," *International Image Sensor Society - 2019 Workshop IISW,* 2019.

[22] J. Ma, D. Starkey, A. Rao, K. Odame and E. R. Fossum, "Characterization of Quanta Image Sensor Pump-Gate Jots With Deep Sub-Electron Read Noise," *Journal of the Electron Devices Society,* vol. 3, no. 6, pp. 472 - 480, 2015.

[23] J. Ma and E. R. Fossum, "A pump-gate jot device with high conversion gain for a quanta image sensor," *IEEE J. Electron Devices Soc.,* vol. 3, no. 2, pp. 73 - 77, 2015.

[24] E. R. Fossum, "Modeling the Performance of Single-Bit and Multi-Bit Quanta Image Sensors," *Journal of the Electron Devices Society,* vol. 1, no. 9, pp. 166 - 174, 2013.

[25] E. Fossum, "Photon Counting Error Rates in Single-Bit and Multi-Bit Quanta Image Sensors," *IEEE Journal of the Electron Devices Society,* 2016.

[26] S. Mims, K. B. Cho, S. Ansari, H. Do, K. Nguyen, W. Tiasn, A. Lopez and S. Vo, "0.3e- Read Noise @30fps 9.5Mpixel CMOS Image Sensor for Scientific Applications Requiring Photon Counting," August 2025. [Online]. Available: https://fairchildimaging.com/hubfs/White%20Papers/HWK4123%20White%20Paper_Aug%202025.pdf?hsLang=en.

[27] N. V. Loukianova, H. O. Folkerts, J. P. V. Maas, D. W. E. Verbugt, A. J. Mierop, W. Hoekstra and E. Roks, "Leakage Current Modeling of Test Structures for Characterization of Dark Current in CMOS Image Sensors," *IEEE Transactions on Electron Devices,* vol. 50, no. 1, pp. 77 - 83, 2003.

[28] D. McGrath, S. Tobin, V. Goiffon, P. Magnan and A. Le Roch, "Dark Current Limiting Mechanisms in CMOS Image Sensors," *Electronic Imaging,* vol. 30, no. 11, pp. 354-1 354-8, 2018.

[29] E. R. Fossum and D. B. Hondongwa, "A Review of the Pinned Photodiode for CCD and CMOS Image Sensors," *IEEE Journal of the Electron Devices Society,* vol. 2, no. 3, pp. 33 - 43, 2014.

[30] A. Boukhayma, A. Peizerat and C. Enz, "Noise Reduction Techniques and Scaling Effects towards Photon Counting CMOS Image Sensors," *Sensors,* vol. 16, no. 4, 2016.

[31] N. A. W. Dutton, I. Gyongy, L. Parmesan and R. K. Henderson, "Single Photon Counting Performance and Noise Analysis of CMOS SPAD-based Image Sensors," *Sensors,* vol. 16, no. 1122, pp. 1-17, 2016.

[32] N. Teranishi, "Required Conditions for Photon-Counting Image Sensors," *IEEE Transactions on Electron Devices,* vol. 59, no. 8, pp. 2199-2205, 2012.

[33] B. Hanold, D. Figer, J. Lee, K. Kolb, I. Marcuson, E. Corrales, J. Getty and L. Mears, "Large format MBE HgCdTe on silicon detector development for astronomy," *SPIE,* p. 96090Y, 2015.

[34] J. R. Janesick, Photon Transfer: DN → λ, Bellingham, WA: SPIE Press, 2007.

[35] K. Cho and B. Johnson, "0.5e- rms Read Noise CMOS Image Sensors and Sub-Electron Image Processing for Night Vision Application," *International Image Sensor Workshop,* 2023.

[36] S. Xie and A. Theuwissen, "Compensation for Process and Temperature Dependency in a CMOS Image Sensor," *Sensors,* vol. 19, no. 4, 2019.

[37] Y. Chen, A. J. Mierop and A. J. P. Theuwissen, "A CMOS Image Sensor With In-Pixel Buried-Channel Source Follower and Optimized Row Selector," *IEEE Transactions on Electron Devices,* vol. 56, no. 11, pp. 2390-2397, 2009.

[38] C. Layden, J. Juneau, G. Pettersson, N. Lourie, B. Schneider, B. LaMarr, F. E. Angile, F. Farag, M. Luo, Z. Z. Ong and G. Furész, "Characterization of the Teledyne COSMOS Camera: A Large Format CMOS Image Sensor for Astronomy," *Journal of Astronomical Telescopes, Instruments, and Systems,* vol. 11, no. 2, p. 026003, 2025.

[39] S. Baggett, "WFC3 TV3 Testing: Quantum Yield in the UV," WFC3 Instrument Science Report, 2008.

[40] L. Werner, U. Linke, I. Müller, T. Kubarsepp, M.-M. Sildoja, T. Tran and J. Gran, "Quantum yield in induced-junction silicon photodiodes at wavelengths around 400 nm," *Metrologia,* vol. 61, no. 3, 2024.

[41] D. F. Figer, B. J. Rauscher, M. W. Regan, E. Morse, J. Balleza, L. Bergeron and H. S. Stockman, "Independent Testing of JWST Detector Prototypes," *SPIE,* vol. 5167, p. 270, 2004.

[42] I. A. Strakhov, B. S. Safonov and D. V. Cheryasov, "Speckle Interferometry with CMOS Detector," *Astrophysical Bulletin,* vol. 78, no. 2, 2023.

[43] M. Regan and L. Bergeron, "Zero Dark Current in H2RG Detectors: it is all Multiplexer Glow," *Journal of Astronomical Telescopes, Instruments, and Systems,* vol. 6, no. 016001.

[44] A. R. Khan, E. Hamden, G. Kyne, A. D. Jewell, J. Henessey, S. Nikzad, V. Picouet, O. Jones, H. Bradley, N. Kerkeser, Z. Lin, B. Parker, G. West, J. Ford and F. Gacon, "Advancing Ultraviolet Detector Technology for future missions: Investigating the dark current plateau in silicon

detectors using photon-counting EMCCDs," in *Proceedings of SPIE, Space Telescopes and Instrumentation*, Yokohama, Japan, 2024.

[45] C. C. Stark, "ExoVista: A Suite of Planetary System Models for Exoplanet Studies," *The Astronomical Journal,* vol. 163, no. 3, p. 11, 2022.

[46] E. Alexani, "Simulation of a Single Photon Counting Photonic Spectrograph for Direct Imaging of Exoplanet Atmospheres with the Habitable Worlds Observatory," *ProQuest,* p. 126, 2024.

[47] E. Alexani, D. Figer and P. Gatkine, "Simulation of a single photon counting photonic spectrograph for exoplanet atmospheric characterization," in *Proceedings of the SPIE*, Yokohama, 2024.

[48] L. Pueyo, C. Stark, R. Juanola-Parramon, N. Zimmerman, M. Bolcar, A. Roberge, G. Arney, G. Ruane, A. J. Riggs, R. Belikov, D. Sirbu, D. Redding, R. Soummer, I. Laginja and S. Will, "The LUVOIR Extreme Coronagraph for Living Planetary Systems (ECLIPS) I: searching and characterizing exoplanetary gems," in *Proceedings of SPIE*, 2019, 2019.

[49] M. Damiano and R. Hu, "Reflected Spectroscopy of small exoplanets II: characterization of terrestrial exoplanets," *The Astronomical Journal,* vol. 163, p. 14, 2022.

[50] S. Seager and D. Deming, "Exoplanet Atmospheres," *Annual Review of Astronomy and Astrophysics,* vol. 48, pp. 631-672, 2010.

[51] E. Schwieterman, N. Kiang, M. Parenteau, C. Harman, S. DasSarma, T. Fisher, G. Arney, H. Hartnett, C. Reinhard, S. Olson, V. Meadows, C. Cockell, S. Walker, J. Grenfell, S. Hedge, S. Rugheimer, R. Hu and T. Lyons, "Exoplanet Biosignatures: A Review of Remotely Detectable Signs of Life," *Astrobiology,* vol. 18, no. 6, p. 46, 2018.

[52] K. Y. Feng, T. D. Robinson, J. J. Fortney, R. E. Lupu, M. S. Marley, N. K. Lewis, B. Macintosh and M. R. Line, "Characterizing Earth Analogs in Reflected Light: Atmopsheric Retrieval Studies for Future Space Telescopes," *The Astronomical Journal,* vol. 155, no. 5, p. 24, 2018.

[53] V. S. Meadows, C. T. Reinhard, G. N. Arney, M. N. Parenteau, E. W. Schwieterman, S. D. Domagal-Goldman, A. P. Lincowski, K. R. Stapelfeldt, H. Rauer, S. DasSarma, S. Hedge and e. a. Narita, "Exoplanet Biosignatures: Understanding Oxygen as a Biosignature in the Context of Its Environment," *Astrobiology,* vol. 18, no. 6, pp. 630-662, 2018.

[54] T. D. Brandt and D. S. Spiegel, "Prospects for Detecting Oxygen, Water, and Chlorophyll on an Exo-Earth," *PNAS,* vol. 111, p. 11, 2014.

[55] B. J. Rauscher and D. Fixsen, "JWST NIRSpec's Cosmic Ray Experience at L2," *Publications of the Astronomical Society of the Pacific,* vol. 137, no. 9, p. 095003, 2025.

[56] L. D. Feinberg and B. Mennesson, "Exploratory Analytic Cases and Coronagraph Exploratory Cases," 3 June 2025. [Online]. Available: https://ntrs.nasa.gov/api/citations/20240006492/downloads/Exploratory%20and%20Analytic%20Cases%20HWO.pdf.

[57] J. Vallerga, O. H. W. Siegmund, A. Tremsin and J. McPhate, "Current and future capabilities of MCP detectors for UV-Vis instruments," September 2011. [Online]. Available:

https://assets.science.nasa.gov/content/dam/science/astro/programs/cosmic-origins/events/2011/stsci_sept2011/Vallerga_COPAG2011.pdf.

[58] L. K. Harding, R. Demers, M. E. Hoenk, P. Peddada, B. Nemati, M. Cherng, D. Michaels, L. S. Neat, A. Loc, N. L. Bush, D. J. Hall, N. J. Murray, J. P. D. Gow, R. Burgon, A. D. Holland and A. L. Reinher, "Technology advancement of the CCD201-20 EMCCD for the WFIRST coronagraph instrument: sensor characterization and radiation damage," *Journal of Astronomical Telescopes, Instruments, and Systems,* vol. 2, no. 1, p. 011007, 2015.

[59] Nüvü Cameras, "nuvucameras_hnu512v3.4," 03 June 2025. [Online]. Available: https://www.nuvucameras.com/wp-content/uploads/nuvucameras_hnu512v3.4.6.pdf.

[60] Teledyne e2v, "CCD201-20 BSI Datasheet (v11).pdf," 22 05 2025. [Online]. Available: https://www.teledynespaceimaging.com/en-us/Products_/Documents/ccd-datasheets/CCD201-20%20BSI%20Datasheet%20(v11).pdf.

[61] A. N. Wilkins, M. W. McElwain, N. T. J., B. J. Rauscher, J. F. Rothe, M. Malatesta, G. M. Hilton, J. R. Bubeck, C. A. Grady and D. J. Lindler, "Characterization of a photon counting EMCCD for space-based high contrast imaging spectroscopy of extrasolar planets," in *Proceedings of SPIE, High Energy, Optical, and Infrared Detectors for Detectors for Astronomy VI*, Montréal, 2014.

[62] Teledyne e2v, "CCD97-00 BSI Datasheet (v10).pdf," 21 05 2025. [Online]. Available: https://www.teledynespaceimaging.com/en-us/Products_/Documents/ccd-datasheets/CCD97-00%20BSI%20Datasheet%20(v10).pdf.

[63] L. Barak, I. M. Bloch, A. Botti, M. Cababie, G. Cancelo, L. Chaplinsky, F. Chierchie, M. Crisler, A. Drlica-Wagner, R. Essig, J. Estrada, E. Etzion, G. Fernandez Moroni and D. Gift, "SENSEI: Characterization of Single-Electron Events Using a Skipper Charge-Coupled Device," *Physical Review Applied,* vol. 17, no. 1, p. 014022, 2022.

[64] E. Marrufo Villalpando, A. Drlica-Wagner, A. A. Plazas Malagon, A. Bakshi, M. Bonati, J. Campa, B. Cancino, C. R. Chavez, J. Estrada, G. Fernandez Moroni, L. Fraga, M. E. Gaido and S. Holland, "Characterization and Optimization of Skipper CCDs for the SOAR Integral Field Spectrograph," *Publications of the Astronomical Society of the Pacific,* vol. 136, no. 4, p. 045001, 2024.

[65] A. Drlica-Wagner, E. Marrufo Villalpando, J. O'Neil, J. Estrada, S. Holland, N. Kurinsky, T. Li, G. Fernandez Moroni, J. Tiffenberg and U. Sho, "Characterization of skipper CCDs for cosmological applications," in *Proceedings of SPIE, X-Ray, Optical, and Infrared Detectors for Astronomy IX*, 2020.

[66] G. Fernández Moroni, J. Estrada, G. Cancelo, S. E. Holland, E. E. Paolini and H. T. Diehl, "Sub-electron readout noise in a Skipper CCD fabricated on high resistivity silicon," *Experimental Astronomy,* vol. 34, pp. 43-64, 2012.

[67] E. R. Fossum, J. Ma, S. Masoodian, L. Anzagira and R. Zizza, "The Quanta Image Sensor: Every Photon Counts," *Sensors,* vol. 16, no. 8, 2016.

[68] G. Hopkinson, D. Purll, A. Abbey, A. Short, D. Watson and A. Wells, "Active pixel array devices in space missions," *Nuclear Instruments and Methods in Physics Research Section A: Accelerators, Spectrometers, Detectors and Associated Equipment,* vol. 513, no. 1, pp. 327-331, 2003.

[69] J. Ma, S. Masoodian, D. A. Starkey and E. R. Fossum, "Photon-Number-Resolving Megapixel Image Sensor at Room Temperature without Avalanche Gain," *Optica,* vol. 4, no. 12, pp. 1474-1481, 2017.

[70] J. P. D. Gow, N. J. Murray, A. D. Holland, D. J. Hall, M. Cropper, D. Burt, G. Hopkinson and L. Duvet, "Assessment of space proton radiation-induced charge transfer inefficiency in the CCD204 for the Euclid space observatory," *Journal of Instrumentation,* vol. 7, no. 1, 2012.

[71] G. R. Hopkinson, A. Mohammadzadeh and R. Harboe-Sorensen, "Radiation effects on a radiation-tolerant CMOS active pixel sensor," *IEEE Transactions on Nuclear Science,* vol. 51, no. 5, pp. 2753-2762, 2004.

[72] R. Massey, "Charge Transfer Inefficiency in the Hubble Space Telescope since Servicing Mission 4," *Monthly Notices of teh Royal Astronomical Society,* vol. 409, no. 1, pp. L109-L113, 2010.

[73] J. Anderson and J. E. & Ryon, "Improving the Pixel-Based CTE-correction Model for ACS/WFC," Instrument Science Report ACS 2018-04, 2018.

[74] R. Massey, C. Stoughton, A. Leauthaud, J. Rhodes, A. Koekemoer, R. Ellis and E. Shaghoulian, "Pixel-based correction for Charge Transfer Inefficiency in the Hubble Space Telescope Advanced Camera for Surveys," *Monthly Notices of the Royal Astronomical Society,* vol. 401, no. 1, pp. 371-384, 2009.

[75] S. Chen, J. Ma, D. B. Hondongwa and E. R. Fossum, "High Conversion-Gain Pinned-Photodiode Pump-Gate Pixels in 180-nm CMOS Process," *IEEE Journal of the Electron Devices Society,* vol. 5, no. 6, pp. 509-517, 2017.

[76] Y. Chen, X. Wang, A. J. Mierop and A. J. P. Theuwissen, "A CMOS Image Sensor With In-Pixel Buried-Channel Source Follower and Optimized Row Selector," *IEEE Transactions on Electron Devices,* vol. 56, no. 11, pp. 2390-2397, 2009.

[77] K. D. Stefanov, M. J. Prest, M. Downing, E. George, N. Bezawada and A. D. Holland, "Simulations and Design of a Single-Photon CMOS Imaging Pixel Using Multiple Non-Destructive Signal Sampling," *Sensors,* vol. 20, no. 7, 2020.

[78] S. Sim and J. Jun, "Advancements in Active-Pixel-Type CMOS Image Sensor Design Techniques and Architectures for Wide Dynamic Range," *Sensors,* vol. 26, no. 2, 2026.

[79] F. Zhang and H. Niu, "A 75-ps Gated CMOS Image Sensor with Low Parasitic Light Sensitivity," *Sensors,* vol. 16, no. 7, 2016.

[80] P. O'Connor, V. Radeka, D. Figer, J. Geary, D. Gilmore, J. Oliver, C. Stubbs, P. Takacs and J. Tyson, "Study of silicon sensor thickness optimization for LSST," *SPIE Astronomical Telescopes + Instrumentation,* vol. 6276, 2006.

[81] S. Nikzad, M. E. Hoenk, F. Greer, B. Jacquot, S. Monacos, T. J. Jones, J. Blacksberg, E. Hamden, D. Schiminovich, C. Martin and P. Morissey, "Delta-doped electron-multiplied CCD with absolute quantum efficiency over 50% in the near to far ultraviolet range for single photon counting applications," *Optica Applied Optics,* vol. 51, no. 3, pp. 365-369, 2012.

[82] M. Hoenk, T. Jones, M. Dickie, F. Greer, T. Cunningham, E. Blazejewski and S. Nikzad, "Delta-doped back-illuminated CMOS imaging arrays: progress and prospects," *SPIE Photonic Devices + Applications,* vol. 7419, 2009.

[83] M. E. Hoenk, A. D. Jewell, G. Kyne, J. Hennessy, T. Jones, S. Cheng, S. Nikzad, D. Morris, K. Lawrie and J. Skottfelt, "2D-doped silicon detectors for UV/optical/NIR and x-ray astronomy," in *Proceedings of SPIE*, 2022.

## Biographies

**Edwin Alexani** is a graduate student at the Rochester Institute of Technology (RIT), pursuing a doctorate in astrophysical sciences and technology (AST). His interests include detector development for astrophysics with a focus on HWO and exoplanets. He is also interested in massive star clusters, Wolf-Rayet (WR) stars, and WR galaxies. He received his MS in AST from RIT in 2024, and his BS in astrophysics from the University of California, Los Angeles (UCLA) in 2022.

**Justin P. Gallagher** is the chief laboratory engineer at the Center for Detectors within the Rochester Institute of Technology. He serves as the project engineer for multiple development programs that aim to advance optical and near-infrared single-photon-counting large-format detector arrays. He received his BS in physics and his MS in astrophysical sciences and technology from RIT in 2020. He designs, develops, and deploys devices to enable scientific discoveries.

**Donald F. Figer** is a professor and director of the Center for Detectors and the Future Photon Initiative at Rochester Institute of Technology. He developed detectors for the Hubble Space Telescope and the James Webb Space Telescope (JWST). He received numerous awards, e.g., the NASA Space Act Award for characterizing competing JWST prototype detectors and the AURA STScI Technology and Innovation Award for founding the Independent Detector Testing Laboratory at Johns Hopkins University.